\documentclass[prb,amsmath,amysymb,amsfonts,superscriptaddress,floatfix,twocolumn]{revtex4-1} 
\usepackage{color}
\usepackage{bm}
\usepackage{graphicx}
\usepackage{hyperref}

\begin{document}
\title{Projected Quasiparticle Theory for Molecular Electronic Structure}
\author{Gustavo E. Scuseria}
\affiliation{Department of Chemistry, Rice University, Houston, Texas 77005, USA}
\affiliation{Department of Physics and Astronomy, Rice University, Houston, Texas 77005, USA}
\author{Carlos A. Jim\'enez-Hoyos}
\affiliation{Department of Chemistry, Rice University, Houston, Texas 77005, USA}
\author{Thomas M. Henderson}
\affiliation{Department of Chemistry, Rice University, Houston, Texas 77005, USA}
\affiliation{Department of Physics and Astronomy, Rice University, Houston, Texas 77005, USA}
\author{Kousik Samanta}
\affiliation{Department of Chemistry, Rice University, Houston, Texas 77005, USA}
\author{Jason K. Ellis}
\affiliation{Department of Chemistry, Rice University, Houston, Texas 77005, USA}
\affiliation{Department of Physics and Astronomy, Rice University, Houston, Texas 77005, USA}
\date{\today}

\begin{abstract}
We derive and implement symmetry-projected Hartree-Fock-Bogoliubov (HFB) equations and apply them to the molecular electronic structure problem. All symmetries (particle number, spin, spatial, and complex conjugation) are deliberately broken and restored in a  self-consistent variation-after-projection approach. We show that the resulting method yields a comprehensive black-box treatment of static correlations with effective one-electron (mean-field) computational cost. The ensuing wave function is of multireference character and permeates the entire Hilbert space of the problem. The energy expression is different from regular HFB theory but remains a functional of an independent quasiparticle density matrix. All reduced density matrices are expressible as an integration of transition density matrices over a gauge grid. We present several proof-of-principle examples demonstrating the compelling power of projected quasiparticle theory for quantum chemistry.
\end{abstract}
\maketitle

\section{Introduction}
Although it has been more than 80 years since the formulation of Schr\"odinger's equation, there is still no black-box computationally efficient treatment of strong correlations in electronic structure theory. Strong correlations, also known as static or non-dynamic correlations, appear from exact and near-degeneracies in the wave function that render the one-determinant Hartree-Fock (HF) picture qualitatively incorrect. By ``computationally efficient'', we mean an effective one-electron theory with mean-field computational cost. Of course, exact diagonalization of the molecular Hamiltonian in the Hilbert space of the problem (a method known as full configuration interaction or FCI) includes all strong correlations but it is impractical except for the smallest systems because of its combinatorial computational cost. Other high accuracy models involve at least $\mathcal{O} (M^6)$ computational effort, where $M$ is the number of orbitals.  The method presented in this paper accomplishes the goal of being an effective one-electron theory with mean-field computational cost, accurately describing strong correlations (and more) of finite systems in a black-box manner.

We derive and implement a wave function method named Projected Quasiparticle Theory (PQT) based on symmetry-projected Hartree-Fock-Bogoliubov (HFB) equations. All symmetries (particle number, spin, spatial, and complex conjugation) are here deliberately broken and restored in a self-consistent approach. We start, that is, with a symmetry broken single Slater determinant, and then restore the symmetry via projection to obtain a multi-determinantal wave function which is variationally superior.  Projection methods on a deformed (\textit{i.e.} broken symmetry) HFB state have been used in nuclear physics for many years.\cite{ring1980,blaizot1985} Unlike those used in quantum chemistry,\cite{lowdin1955} these projections are based on the generator coordinate method.\cite{ring1980}  In particular, number projected Hartree-Fock-Bogoliubov (PHFB) is perhaps most widely used because the attractive character of the nucleon-nucleon interaction usually leads to HFB solutions that have lower energy than HF. The work that we here present builds upon the number projection formalism of Sheikh and Ring.\cite{sheikh2000,sheikh2001,sheikh2002} We expand and develop their technique to include many molecular symmetries not previously considered.

PHFB is a variational problem where a deformed quasiparticle determinant $\left| \Phi \right\rangle$ is optimized in the presence of projection operators $\hat{P}$ which can represent a collection of symmetries.  Sheikh and Ring\cite{sheikh2000} realized that the energy arising from this minimization (see below) can be expressed as a functional of the HFB regular density matrix and anomalous density matrix. Thus, this problem can be solved by diagonalization of an effective Hamiltonian matrix. Here, we extend their work and consider projections onto eigenfunctions of the particle number, spin rotation (both $S^2$ and $S_z$), complex conjugation, and spatial symmetry operators. In essence, we here present a method where all symmetries in the molecular wave function\cite{fukutome1981,stuber2003} may be \textit{deliberately} broken and variationally restored. This includes both continuous (infinite dimension) such as spin rotation and discrete group representations such as spatial symmetry and complex conjugation; the latter two had not been considered previously. The resulting mathematical problem of variation-after-projection (VAP) seems formidable at first glance, yet as we show below, it can be converted into essentially one of diagonalization of an effective one-quasiparticle Hamiltonian with mean-field computational cost.  Alternatives to the diagonalization approach that we  pursue are discussed in the nuclear physics literature.\cite{bender2003,schmid2004}

The research presented here was motivated by recent work in our group. In a series of papers\cite{tsuchimochi2009,scuseria2009,tsuchimochi2010,tsuchimochi2010b,TripletPairing}, we have proposed Constrained-Pairing Mean-Field Theory (CPMFT), a method for dealing with strong correlations.  By introducing a fictitious attractive pairing interaction between electrons in an active space, CPMFT accounts for strong correlations and properly dissociates molecules into fragments with correct spatial and spin symmetries. CPMFT is black-box but does not have a wave function; it is in practice a one-particle density matrix functional whose two-particle density matrix is not N-representable but whose energy can be optimized via an effective HFB diagonalization problem.\cite{staroverov2002}  HFB breaks particle number symmetry, so CPMFT requires a chemical potential for controlling electron number.

In its singlet-paired version,\cite{TripletPairing} CPMFT is intimately connected with Unrestricted HF (UHF).\cite{tsuchimochi2010b} Its Lagrange multipliers for constraining pairing interactions within an active space inspired the resolution of an old problem: how to obtain the ROHF wave function within an UHF framework.\cite{tsuchimochi2010c}  This is essentially a constrained variation problem of spin-projection of an unrestricted determinant.  We should note that in open-shell cases, spin symmetry breaking occurs spontaneously when using unrestricted orbitals.  On the other hand, when dealing with singlet states, it does not matter if the restricted HF (RHF) wave function is stable or unstable; in both cases the VAP wave function originating from a UHF guess becomes multireference in nature. This is akin to ``deliberate'' symmetry breaking followed by a restoration step. In other words, spontaneous symmetry breaking (due to so-called ``HF instabilities'') is not a necessary condition in VAP schemes. Curiously, CPMFT can be interpreted as deliberate number symmetry breaking (triggered by HFB with a fictitious attractive interaction), followed by a symmetry restoration step generated by its energy and two-particle density matrix definition.\cite{tsuchimochi2009,scuseria2009,tsuchimochi2010,tsuchimochi2010b}

The lessons learned with CPMFT and our desire to develop a \textit{wave function} theory for strong correlations, steered us into projection schemes. Number projected HFB (also known as projected Bardeen-Cooper-Schrieffer or PBCS) is widely discussed in textbooks\cite{ring1980,blaizot1985} and yields the Antisymmetrized Geminal Power (AGP) wave function of quantum chemistry, a model that has been extensively studied.\cite{linderberg1980,sangfelt1981,ortiz1981,kurtz1981,coleman1997,mazziotti2000,coleman2000,weiner2002,staroverov2002,casula2004,bajdich2006}  As shown here, projected HFB can be used not only to optimize AGP wave functions with singlet geminals (singlet pairing) but also to optimize broken symmetry geminals which, to the best of our knowledge, has not been done before.  When all symmetries are broken, the resulting projected wave function is of multireference character, one that permeates the entire Hilbert space of the problem.  By this, we mean that every Slater determinant built from the natural orbitals of our broken symmetry determinant is allowed to overlap with our projected wave function.  Remarkably, the projected wave function accomplishes this feat with a linear number of parameters in an implicit factorization scheme driven by the variational principle.

Viewed in their natural orbital basis, closed-shell singlet-paired AGP wave functions belong to the seniority zero sector of the Hilbert space of the problem and suffer from known drawbacks.\cite{staroverov2002}  Seniority is here defined as twice the number of broken electron pairs in a determinant (\textit{i.e.}, the number of singly occupied spatial orbitals).  For more details about the seniority concept and its use in electronic structure theory, see Ref. \onlinecite{bytautas2011}.  Spin projection on top of number projection forces the wave function to permeate to all seniority sectors. By deliberatively breaking spin symmetry and forcing $\alpha$ and $\beta$ spatial orbitals to be different, all electron pairs are broken and spin contamination is introduced. In our case, however, spin projection operators in the HFB context restore the correct quantum numbers for UHF ($S^2$) and Generalized HF (GHF) ($S^2$ and $S_z$) orbitals and density matrices. In GHF the spin orbitals are linear combinations of $\alpha$ and $\beta$ spin, yielding a density matrix with all spin blocks populated (usually referred to as noncollinear configuration).  Access to the triplet pairing channel is required if spin symmetry is to be broken and the broken symmetry wave function has HFB character.\cite{TripletPairing}  We note in passing that the spin projection method developed in this work provides a solution to the extended Hartree-Fock method of L\"owdin and Mayer\cite{mayer1980} and Goddard's GF method\cite{goddard1967}, albeit accomplished simultaneously with number projection.

Both number and spin projection operators (or their generalization in the case of non-singlet spin) can be written as integrals over gauge angles in the generator coordinate approach that we employ.\cite{ring1980,blaizot1985} The restoration of these symmetries is mathematically accomplished here via discretization of the gauge integral over modest size grids.  The corresponding symmetries are U(1) for number and SU(2) [homeomorphic to SO(3)] for triaxial noncollinear spin-projection.

Restoration of discrete symmetries like point group (space) and complex conjugation yield wave functions that are linear combinations of the generalized AGP wave functions discussed above. Complex conjugation is an antiunitary symmetry\cite{wigner1960} and its restoration requires special consideration, as discussed below.

The effective one-electron PHFB Hamiltonian includes both Fock and pairing pieces that depend on transition density matrices that are defined over the gauge grid. These transition density matrices yield via gauge integration a correlated and factorizable two-particle density matrix. Computationally, the most expensive step of PHFB requires that the two-electron integrals be contracted with the transition density matrices at every grid point, so the computational scaling of this generalized form of PHFB is comparable to the mean-field cost of single reference HF times the number of grid points, which we show below is fairly insensitive to the size of the system.  Depending on basis set, system size, and energy gap, this computational cost\cite{strout1995} will be $\mathcal{O}(M^2)$ to $\mathcal{O}(M^3)$.  The entire myriad of linear scaling tools (including alternatives to diagonalization) is of course available to help further reduce this computational cost to $\mathcal{O}(M)$.\cite{scuseria1999}  The integration over grid points is trivially parallelizable.

Remarkably, the PHFB wave function presented in this work can be obtained with mean-field computational cost even though its multireference character permeates the entire FCI space. The fundamental reason behind this feat resides in a projected energy expression that is of HFB form and can be solved for by a standard HFB diagonalization. 

From a mathematical perspective, the results here obtained are supported by an implicit coherent state representation of non-orthogonal wave functions out of which the component of desired symmetry can be projected while simultaneously satisfying the variational principle.\cite{ring1980,blaizot1985}  The class of multireference wave functions studied in this work, which can be written as a quasiparticle density matrix functional, have remained completely unexplored in electronic structure theory.  The only previously known density matrix functional that has an accompanying wave function is Hartree-Fock.

The remainder of this paper is organized as follows. We discuss the main equations of our method, deferring details to the Appendix.  A number of benchmark results are then presented and discussed followed by concluding remarks.

\section{Theory}
\subsection{General Considerations}
Fundamentally, our multireference wave functions are simply the result of projection from a broken symmetry quasiparticle determinant.  We will provide a few necessary details of quasiparticle theory here; the interested reader should consult one of the many textbooks available on the subject.\cite{ring1980,blaizot1985}  The quasiparticle creation operators that we use are obtained by a Bogoliubov-De Gennes transformation of the standard electron creation and annihilation operators:
\begin{equation}
\beta_i^\dagger = \sum ( U_{ji} a_j^\dagger + V_{ji} a_j).
\end{equation}
A quasiparticle determinant can then be written as
\begin{equation}
|\Phi\rangle = \prod_{i=1}^{M/2} \beta_i |\rangle
\end{equation}
where $M$ is the dimension of the single-particle basis and $|\rangle$ is the empty state.  Note that this quasiparticle determinant dwells in Fock space rather than in Hilbert space.  In the Hartree-Fock-Bogoliubov method, one variationally minimizes the expectation value of the Hamiltonian with respect to the coefficients $\mathbf{U}$ and $\mathbf{V}$ defining the quasiparticle orbitals.  Given the coefficients $\mathbf{U}$ and $\mathbf{V}$, one can form the Hermitian density matrix
\begin{equation}
\bm{\rho} = \mathbf{V}^\star \, \mathbf{V}^\mathsf{T}
\end{equation}
and the antisymmetric anomalous density matrix
\begin{equation}
\bm{\kappa} = \mathbf{V}^\star \, \mathbf{U}^\mathsf{T} = -\mathbf{U} \, \mathbf{V}^\dagger.
\end{equation}
We remind the reader that the \textit{unprojected} HFB wave function converges to HF in cases where the two-body interaction is repulsive, as in electronic structure.

It may be instructive to write the HFB determinant in a slightly different manner.  In the so-called HFB ``canonical'' (natural orbital) basis, the quasiparticle determinant is
\begin{equation}
| \Phi \rangle = \mathcal{N} \prod_{k=1}^s \left( 1 + \zeta_k  a_k^{\dagger} a_{\bar{k}}^{\dagger} \right) | \rangle,
\label{Eqn:AGP}
\end{equation}
where $\mathcal{N}$ is a normalization factor and $\zeta_k = v_k / u_k$.  The product runs over the $s$ orbitals defining the subspace over which the HFB wave function is allowed.  Usually, $s = M/2$, but this is not always the case.  Indeed, if $s = N/2$ where $N$ is the number of electrons, then the number projected HFB corresponds exactly to HF.  Note that $s$ is known in the AGP literature as the \textit{rank} of the geminal.  In writing Eqn. \ref{Eqn:AGP}, we have used the indices $k$ and $\bar{k}$ to represent the orbitals that are paired.  Typically the paired orbitals are chosen to be $k \alpha$ and $k \beta$, the two spin orbitals formed from the same spatial orbital.  When this is so, the projected HFB wave function is manifestly of seniority zero.  However, one can choose to pair orbitals differently, a point to which we shall return later.

From the form of Eqn. \ref{Eqn:AGP} it is clear that the HFB determinant contains contributions from states of many different particle numbers, and that for each particle number, number projection of HFB yields a linear combination of many determinants.  In fact, when $s=M/2$ the number projected unrestricted HFB yields a linear combination of \textit{every} determinant, which is what we mean when we say that the projected HFB wave function permeates Hilbert space.  Pairing the spatial orbitals $k \alpha$ and $k \beta$ instead yields a projected HFB wave function which is a linear combination of every determinant of seniority zero.  In either case, each determinant has its own coefficient, but those coefficients are determined by the HFB parameters $\zeta_k$.  In other words the coefficients of the different determinants are factorized.

Approximate wave functions such as quasiparticle determinants need not possess the same symmetries as the exact solutions. In fact, constraining the variationally optimized state to preserve certain symmetries can only lead to higher energy solutions as this reduces the variational manifold, a fact now known as L\"owdin's \textit{symmetry dilemma}.\cite{lykos1963}

Projection operators can be used to restore symmetries of the Hamiltonian from a broken-symmetry approximate wave function, thus avoiding the symmetry dilemma. One can use the projection operators in two ways:
\begin{itemize}
\item In the projection-after-variation (PAV) scheme, one optimizes a broken symmetry wave function, and then performs a single-shot symmetry restoration with the projection operator.
\item In the variation-after-projection (VAP) scheme, the optimization is carried out in the presence of the projection operator. That is, one minimizes the expectation value of the projected wave function with respect to variations of the underlying deformed state.
\end{itemize}
In most cases VAP is preferred since it uses all the variational flexibility available and avoids the artifactual phase transitions associated with symmetry breaking of the underlying wave function, which may become more pronounced when using the PAV scheme.\cite{schlegel1986,GHF}  Nonetheless, carrying out the full variational optimization with the VAP scheme may lead to equations that become significantly more involved than solving the original variational problem.\cite{mayer1980,goddard1967} In finite systems, quantum fluctuations incorporated in the VAP scheme lead to multireference wave functions that remove all artifactual phase transitions due to spontaneous symmetry breaking. This will be clear in the dissociation curves presented below.  We note in passing that in the thermodynamic limit (\textit{infinite} systems), VAP and PAV based on a mean-field state do not improve the energy over that of the broken symmetry mean-field solution.\cite{anderson1984,blaizot1985}  The focus of this paper is on finite systems only.

In the special case when the deformed state is a quasiparticle determinant, it turns out that the VAP equations, though involved, can be solved in a straightforward manner. The nuclear physics literature contains a large number of \textit{approximate} solutions.\cite{ring1980} Exact solutions to certain specific VAP problems in nuclear physics have also been discussed\cite{sheikh2000,schmid2004,bender2003} but not always fully implemented.  The projection operators employed here are based on the generator coordinate method\cite{ring1980,peierls1957,bayman1960,percus1962} and are very different from the traditional approach in quantum chemistry due to L\"owdin.\cite{lowdin1955}   We note that the ``Variation After Mean field Projection In Realistic model spaces'' of Schmid and collaborators share common features with out formalism, though the authors do not optimize the projected HFB state via diagonalization of a Hermitian matrix as we do.\cite{schmid2004}  To the best of our knowledge, projection methods have been applied to the electronic problem only twice,\cite{fernandez2003,schmid2005} and then only with model Hamiltonians.

Given a Hermitian operator $\hat{\Lambda}$ which commutes with the Hamiltonian, we can choose eigenfunctions of an approximate Hamiltonian such as HF or HFB to also be eigenfunctions of $\hat{\Lambda}$.  If we break $\hat{\Lambda}$-symmetry (that is, if our wave function is chosen to not be an eigenfunction of $\hat{\Lambda}$), we can construct a corresponding unitary operator $\hat{U} = \exp{\mathrm{i} \phi (\hat{\Lambda} - \lambda)}$ with parameters $\phi$ and $\lambda$ where $\phi$ is a gauge angle over which we will integrate and $\lambda$ is the desired eigenvalue of $\hat{\Lambda}$.  Integration over $\phi$ will then yield a projection operator $\hat{P}$ which projects eigenstates of $\hat{\Lambda}$ with eigenvalue $\lambda$ out of the the symmetry-broken wave function, provided that all eigenvalues of $\hat{\Lambda}$ are rational.  We sketch the main ideas below for particle number projection as an example.

\subsection{Particle Number Projection as an Example}
Suppose that $|\Phi\rangle$ is a quasiparticle determinant.  As a quasiparticle determinant, it is completely specified by its density matrix $\bm{\rho}$ and anomalous density matrix $\bm{\kappa}$, and is a linear combination of particle number eigenstates:
\begin{subequations}
\begin{align}
|\Phi\rangle &= \sum c_k |\Psi_k\rangle,
\\
\hat{N} |\Psi_k\rangle &= N_k |\Psi_k\rangle,
\end{align}
\end{subequations}
where $\hat{N} = \sum a_i^\dagger a_i$ is the number operator.  We emphasize that the number eigenstates $|\Psi_k\rangle$ are \textit{multideterminantal wave functions} rather than being single determinants.

Then consider the unitary operator $\hat{U}(\theta) = \exp(\mathrm{i} \, \theta \, \hat{N})$.  Its action on $|\Phi\rangle$ is to produce other quasiparticle determinants:
\begin{equation}
\hat{U}(\theta) |\Phi\rangle = \sum c_k \, \mathrm{e}^{\mathrm{i} \, \theta \, N_k} |\Psi_k\rangle \equiv |\Phi(\theta)\rangle.
\label{UnitaryOperator}
\end{equation}
We can project out the component $|\Psi_j\rangle$ with particle number $N_j$ by multiplying by the weight function $w_j(\theta) = \exp(-\mathrm{i} \, \theta \, N_j)/(2\pi)$ and integrating.  That is,
\begin{equation}
c_j |\Psi_j \rangle = \int\limits_0^{2\pi} \, \mathrm{d}\theta \, w_j(\theta) \, \hat{U}(\theta) |\Phi\rangle.
\end{equation}
We will write
\begin{equation}
\hat{P}_j = \int\limits_0^{2\pi} \, \mathrm{d}\theta \, w_j(\theta) \, \hat{U}(\theta),
\label{ProjectionOperator}
\end{equation}
which is the projection operator onto the number eigenstate of interest.

Now, the energy we wish to minimize is simply
\begin{equation}
E = \frac{\langle \Phi| \hat{P}_j^\dagger \, \hat{H} \, \hat{P}_j |\Phi\rangle}{\langle \Phi | \hat{P}_j^\dagger \hat{P}_j |\Phi\rangle}
  = \frac{\langle \Phi| \hat{H} \, \hat{P}_j |\Phi\rangle}{\langle \Phi | \hat{P}_j |\Phi\rangle}
\end{equation}
where we have used the fact that $\hat{P}_j$ is Hermitian, idempotent, and commutes with the Hamiltonian.  The matrix elements in the numerator and denominator can be, in principle, expressed entirely in terms of the projection operator and the density matrices $\bm{\rho}$ and $\bm{\kappa}$ associated with $|\Phi\rangle$.  We then simply minimize $E$ with respect to $\bm{\rho}$ and $\bm{\kappa}$ subject to the constraint that they are compatible with a quasiparticle determinant, and the result defines the VAP energy and wave function.  Projections onto eigenstates of other operators follow a qualitatively similar pattern, though with significant differences in the details.

In general, given a Hermitian constant of motion (symmetry), it is always possible to construct a unitary operator like that in Eqn. \ref{UnitaryOperator}.  This operator creates a manifold of degenerate states from which the desired state of interest can be extracted via a projection operator of the form in Eqn. \ref{ProjectionOperator}.  It is not the purpose of this paper to discuss these mathematical tools in detail. The interested reader is referred to one of the textbooks in the field.\cite{ring1980,blaizot1985}

\subsection{Projection Operators}
Here we wish to briefly review the projection operators we will use.  We wish to take a fairly general form, to make what follows as transparent as possible.

When the generators in which we are interested are continuous, the projection operators we consider can be expressed as 
\begin{equation}
\hat{P}_j(\hat{J}) = \int \, \mathrm{d}\theta \, w_{j,\hat{J}}(\theta) \, \hat{R}(\theta,\hat{J})
\label{Eqn:ProjOperator}
\end{equation}
where we wish to project onto the $j^\textrm{th}$ eigenstate of the operator $\hat{J}$.  The precise form of the weight $w_{j,\hat{J}}(\theta)$ depends on the operator $\hat{J}$ and also on the eigenstate $j$.  The rotation operator $\hat{R}(\theta,\hat{J})$ similarly depends on the symmetry operator $\hat{J}$.  If the generator is instead discrete, the integration is replaced by a summation.

The rotation operator $\hat{R}(\theta,\hat{J})$ acts on a deformed or broken symmetry state in such a way that
\begin{subequations}
\begin{align}
\left|\langle \Phi| \Phi\rangle\right| &= \left|\langle \Phi(\theta,\hat{J}) | \Phi(\theta,\hat{J})\rangle \right|
\\
\left|\langle \Phi| \hat{A} | \Phi\rangle\right| &= \left|\langle \Phi(\theta,\hat{J}) | \hat{A} | \Phi(\theta,\hat{J})\rangle \right|
\end{align}
\end{subequations}
where $|\Phi(\theta,\hat{J})\rangle = \hat{R}(\theta,\hat{J}) | \Phi\rangle$ and where $\hat{A}$ commutes with $\hat{J}$ and thus with $\hat{R}(\theta,\hat{J})$.  In other words, $\hat{R}(\theta,\hat{J})$ acts on the deformed state, preserving the norm and matrix elements of commuting observables up to an overall phase factor.  Typically, $\hat{R}(\theta,\hat{J})$ will be unitary, but it may instead be antiunitary.  We point out that the rotated states form a nonorthogonal set that can be overcomplete, in the same manner as are coherent states.  In fact, as we have already noted, the success of this approach is based on an underlying coherent state representation.\cite{ring1980,blaizot1985}

In this paper we will consider four types of projection operators associated with particle number, spin rotation, point group symmetry, and complex conjugation. We will briefly discuss the form of each of these projection operators.  We will also give the matrix representations $\mathbf{R}(\theta,\hat{J})$ of the rotation operators $\hat{R}(\theta,\hat{J})$ in an orthonormal basis of spin orbitals $|i \sigma\rangle$, where we have written them in spin blocks.  That is,
\begin{equation}
R_{ij}(\theta,\hat{J}) = \langle i | \hat{R}(\theta,\hat{J}) | j \rangle.
\end{equation}

\textit{Number --}
The projection operator associated with particle-number restoration is particularly simple. It is given as an integration over the gauge angle $\phi$, in the form of Eqn. \ref{Eqn:ProjOperator}, with
\begin{align}
w_N(\phi) &= \frac{1}{2 \pi} e^{-\mathrm{i} \phi N},
\\
\hat{R}(\phi,\hat{N}) &= e^{\mathrm{i} \phi \hat{N}},
\\
\mathbf{R}(\phi,\hat{N})
   &= \begin{pmatrix}  \mathrm{e}^{\mathrm{i} \phi} \, \bm{1}  &  \bm{0}   \\
                       \bm{0}                                 &  \mathrm{e}^{\mathrm{i} \phi} \, \bm{1} \end{pmatrix}
\end{align}
where $\hat{N}$ is the number operator and $N$ is the number of particles that the projected wave function shall possess.

\textit{Spin --}
The electronic Hamiltonian is spin free, and thus the wave function should be invariant to rotations of spin (\textit{i.e.} spin can be quantized along an arbitrary axis).  Formally, this procedure corresponds to using a projection operator of the form
\begin{equation}
\hat{P}_S = \sum_{M,K} c_M c_K^\star |S;M\rangle \langle S;K|
\end{equation}
where $S$ refers to the eigenvalue of $\hat{S}^2$, $M$ to the eigenvalue of $\hat{S}_z$, and the $c_M$ are variationally optimized coefficients.  Because we do not generally know the spin eigenfunctions $|S;M\rangle$, we instead write the projector as an integration over spin rotations characterized by the Euler angles $\Omega = (\alpha, \beta, \gamma)$.  That is, we have
\begin{equation}
\hat{P}_S = \sum c_M c_K^\star \int \mathrm{d}\Omega \, w_{S,M,K}(\Omega) \, \hat{R}(\Omega,\hat{S})
\end{equation}
where the weights and the rotation operator are given by by\cite{percus1962,mizusaki2004}
\begin{align}
w_{S,M,K}(\Omega) &= \frac{2S + 1}{8 \pi^2} D^{S \star}_{MK}(\Omega),
\\
\hat{R}(\Omega,\hat{S}) &= e^{\mathrm{i} \alpha \hat{S}_z} \, e^{\mathrm{i} \beta \hat{S}_y} \, e^{\mathrm{i} \gamma \hat{S}_z},
\\
\mathbf{R}(\Omega,\hat{S}) &= \mathbf{R}(\alpha,\hat{S}_z) \,  \mathbf{R}(\beta,\hat{S}_y)  \, \mathbf{R}(\gamma,\hat{S}_z),
\\
\mathbf{R}(\alpha,\hat{S}_z)
   &= \begin{pmatrix}  \mathrm{e}^{\mathrm{i} \alpha/2} \, \bm{1}  &  \bm{0}  \\
                       \bm{0}                                   &  \mathrm{e}^{-\mathrm{i} \alpha/2} \, \bm{1} \end{pmatrix}
\\
\mathbf{R}(\beta,\hat{S}_y)
   &= \begin{pmatrix}  \cos(\beta/2)\, \bm{1}   &  \sin(\beta/2)\, \bm{1}   \\
                      -\sin(\beta/2)\, \bm{1}   &  \cos(\beta/2)\, \bm{1}   \end{pmatrix}.
\end{align}
Here, $D^S_{MK}(\Omega) = \langle S; M | \hat{R}(\Omega,\hat{S}) | S; K\rangle$ are Wigner rotation matrices.\cite{biedenharn1981}

We note that the spin projection operator has the same form as the angular momentum projection operators commonly used in nuclear physics to restore spatial rotations, as the SU(2) algebra characterizing spin is homeomorphic to the SO(3) algebra characterizing angular momentum.  We refer the reader to the exposition of the angular momentum projection operators in Ref. \onlinecite{ring1980}.  Note that if the underlying reference state is an eigenfunction of $\hat{S}_z$ (a collinear state), as is usually the case, then the integrations over $\alpha$ and $\gamma$ are trivial and the projection operator becomes
\begin{equation}
\hat{P}_{S, N_S, N_S} = \frac{2 S + 1}{2} \int_0^\pi \mathrm{d}\beta \, \sin(\beta) \, d^S_{N_S N_S}(\beta) e^{\mathrm{i} \beta \hat{S}_y},
\end{equation}
where $N_S = N_{\alpha} - N_{\beta}$ and $d^s_{N_S N_S}(\beta) = \langle S; N_S | \hat{R}(\beta,\hat{S}_y) | S; N_S\rangle$ is Wigner's small d-matrix.\cite{biedenharn1981}  Note also that in restoring invariance to the axis of spin quantization we also force the wave function to be an eigenfunction of $\hat{S}^2$. This form of the spin projection operator was proposed by Percus and Rotenberg\cite{percus1962} in 1962 and was used in chemistry by Lefebvre and Prat\cite{lefebvre1969} in the context of HF, though their methods and techniques never became widely used.

\textit{Point Group --}
Restoring point group symmetry can follow the same mathematics as used above but is more simply accomplished by diagonalization of the Hamiltonian matrix in the basis of the wave functions $\hat{\mu} \left| \Phi \right\rangle$, where $\hat{\mu}$ are the elements of the point group under consideration. In other words, the projected energy becomes
\begin{equation}
E_j = \frac{\sum_{\mu, \nu} c^\star_{\nu} c_\mu \left\langle \Phi \left| \hat{\nu}^\dagger \hat{H} \hat{\mu}\right| \Phi \right\rangle}
           {\sum_{\mu, \nu} c^\star_{\nu} c_\mu \left\langle \Phi \left| \hat{\nu}^\dagger \hat{\mu} \right| \Phi \right\rangle},
\end{equation}
where $c_\mu$ are variational coefficients.  This is similar to the way other authors restore parity in nuclei.\cite{schmid2004}  The rotation matrices for spatial symmetry restoration correspond to the matrix representation of the symmetry elements of the point group.  

Spatial symmetry breaking and restoration is useful for exploring regions of potential energy surfaces where some spatial symmetry is preserved, and only those symmetry operations which are preserved should be restored in the manner described above.  If different symmetry operators are restored at different nuclear geometries, the resulting potential energy surface is likely to be discontinuous.  One could try to extend our methods by using all the symmetry operators defined at high symmetry points in the diagonalization above, even at points of lower symmetry, but this is not symmetry breaking and restoration and its discussion is beyond the scope of this paper.

\textit{Complex Conjugation --}
The exact wave function in quantum chemistry can be chosen to be an eigenfunction of the complex conjugation operator $\hat{K}$ as long as the Hamiltonian commutes with it, which is always true in the nonrelativistic case where $\hat{H}$ is real.  Complex conjugation is an antiunitary operator so its spectrum does not carry good quantum numbers with it.\cite{wigner1960}  A wave function $|\Psi\rangle = \sum c_i |\Phi_i\rangle$ is an eigenfunction of $\hat{K}$ when
\begin{equation}
\hat{K} |\Psi\rangle = \sum c_i^\star |\Phi_i\rangle = \mathrm{e}^{\mathrm{i} \phi} |\Psi\rangle
\end{equation}
implying
\begin{equation}
\langle \Psi | \hat{K} | \Psi \rangle = \mathrm{e}^{\mathrm{i} \phi}
\label{Eqn:KNorm}
\end{equation}
for some real number $\phi$.  An HFB wave function need not be an eigenfunction of $\hat{K}$, and if the HFB wave function does not satisfy the property of Eqn. \ref{Eqn:KNorm}, then we can diagonalize the Hamiltonian in the basis $\{|\Psi\rangle, \hat{K}|\Psi\rangle\}$.  The resulting wave function has lower energy.

A cautionary note is in order.  The matrix elements involved in the CI problem required to restore point group symmetry can be found by building the rotation matrix corresponding to the operator $\hat{\mu}$ and using the formulas provided below. On the other hand, the matrix elements involving the complex conjugation operator are slightly more involved as there is no associated rotation operator. 

Henceforth, we will suppress dependence on $\hat{J}$ for brevity of notation.

\subsection{Variation-After-Projection Scheme}
In the variation-after-projection scheme, our aim shall be to minimize the expectation value
\begin{equation}
E_j 
  = \frac{\langle \Phi| \hat{H} \hat{P}_j | \Phi\rangle}{\langle \Phi| \hat{P}_j | \Phi \rangle} 
  = \frac{\int \mathrm{d}\theta \, w_j(\theta) \langle \Phi| \hat{H} \hat{R}(\theta) | \Phi \rangle}{\int \mathrm{d}\theta \, w_j(\theta) \langle \Phi | \hat{R}(\theta) | \Phi \rangle}.
\label{Eqn:EPHFB}
\end{equation}
The minimization should be carried over all possible states $\left| \Phi \right\rangle$ of the HFB form.

It is most straightforward to evaluate Hamiltonian elements between quasiparticle determinants in intermediate normalization.  We can define such determinants as
\begin{equation}
|\theta \rangle = \frac{\hat{R}(\theta) | \Phi \rangle}{\langle \Phi | \hat{R}(\theta) | \Phi \rangle}.
\end{equation}
Then the energy expression is
\begin{subequations}
\begin{align}
E_j 
  &= \frac{\int \mathrm{d}\theta \, w_j(\theta) \langle \Phi | \hat{H} \hat{R}(\theta) | \Phi \rangle}{\int \mathrm{d}\theta \, w_j(\theta) \langle \Phi | \hat{R}(\theta) | \Phi \rangle}
\\
  &=  \frac{\int \mathrm{d}\theta \, w_j(\theta)  \, \langle \Phi | \hat{R}(\theta) | \Phi \rangle \, \langle \Phi | \hat{H} |\theta \rangle}{\int \mathrm{d}\theta \, w_j(\theta) \langle \Phi | \hat{R}(\theta) | \Phi \rangle}.
\end{align}
\end{subequations}
Following Sheikh and Ring\cite{sheikh2000}, we make this expression more compact by absorbing the overlap matrix elements into new weight factors, defining
\begin{equation}
y_j(\theta) = \frac{ w_j(\theta)  \, \langle \Phi | \hat{R}(\theta) | \Phi \rangle}{\int \mathrm{d}\theta \,  w_j(\theta)  \, \langle \Phi | \hat{R}(\theta) | \Phi \rangle}
\end{equation}
in terms of which the energy is simply
\begin{equation}
E_j = \int \mathrm{d}\theta \, y_j(\theta) \langle \Phi | \hat{H} | \theta \rangle.
\end{equation}

\subsection{Evaluation of Matrix Elements}
The remaining task is to evaluate the Hamiltonian and overlap elements.  The overlap matrix elements can be expressed as\cite{onishi1968,ring1980}
\begin{equation}
\begin{split}
\langle \Phi | \hat{R}(\theta) | \Phi \rangle &= \exp\left\{\frac{1}{2} \mathrm{Tr}\left[\log\left(\bm{1} - \bm{\mathcal{M}}(\theta)\right)\right]\right\}
\\
 &- \exp\left\{\frac{1}{2} \mathrm{Tr}\left[\log\left(\bm{1} - \bm{\mathcal{M}}(0)\right)\right]\right\}.
\end{split}
\end{equation}
Here and in what follows, we use the matrix
\begin{equation}
\bm{\mathcal{M}}(\theta) = \mathbf{Z}^\star \mathbf{R}(\theta) \mathbf{Z} \mathbf{R}^\mathsf{T}(\theta),
\end{equation}
where $\mathbf{Z} = \mathbf{V}^\star (\mathbf{U}^\star)^{-1}$ is the Thouless matrix\cite{thouless1960,ring1980} characterizing the HFB state $|\Phi\rangle$ in terms of its orbital coefficient matrices $\mathbf{U}$ and $\mathbf{V}$.  We point out here that we have considered only single-particle operators, which places restrictions on the symmetries to be projected.  Note also that we will frequently write $\mathbf{R}_\theta$ in place of $\mathbf{R}(\theta)$ in order to reduce clutter, and so on for other $\theta$-dependent matrices.

The Hamiltonian matrix elements $\langle \Phi | \hat{H} | \theta \rangle = H_\theta$ are
\begin{equation}
\begin{split}
H_\theta
  &= \sum h_{ik} \rho_{ki}(\theta) + \frac{1}{2} \sum \langle ij \| kl \rangle \rho_{ki}(\theta) \rho_{lj}(\theta)
\\
  &+ \frac{1}{4} \sum \langle ij \| kl \rangle \bar{\kappa}^\star_{ij}(\theta) \kappa_{kl}(\theta).
\end{split}
\end{equation}
Here, $h_{ij}$ are the usual one-electron integrals and $\langle ij \|kl \rangle$ are antisymmetrized two-electron integrals in Dirac notation, while $\bm{\rho}_\theta$, $\bm{\kappa}_\theta$, and $\bar{\bm{\kappa}}_\theta$ are transition density matrices given by
\begin{subequations}
\begin{align}
\rho_{ij}(\theta) &= \langle \Phi | a_j^\dagger a_i | \theta \rangle 
\\
                  &= \left[-(\mathbf{Z}^\star)^{-1} \bm{\mathcal{M}}_\theta \left(\bm{1} - \bm{\mathcal{M}}_\theta\right)^{-1} \mathbf{Z}^\star\right]_{ij},
\\
\kappa_{ij}(\theta) &= \langle \Phi | a_j a_i | \theta \rangle
\\
                   & = \left[ (\mathbf{Z}^\star)^{-1} \bm{\mathcal{M}}_\theta\left(\bm{1} - \bm{\mathcal{M}}_\theta\right)^{-1}\right]_{ij},
\\
-\bar{\kappa}_{ij}^\star(\theta) &= \langle \Phi | a_j^\dagger a_i^\dagger | \theta \rangle
\\
                                &= \left[-\left(\bm{1} - \bm{\mathcal{M}}_\theta\right)^{-1} \mathbf{Z}^\star\right]_{ij}
\end{align}
\end{subequations}
One can simplify the Hamiltonian matrix elements as
\begin{equation}
\begin{split}
H_\theta
   &= \frac{1}{2} \mathrm{Tr}\left[(\bm{h} + \bm{\mathcal{F}}_\theta) \bm{\rho}_\theta - \bar{\bm{\Delta}}^\star_\theta \bm{\kappa}_\theta\right]
\\
   &= \frac{1}{2} \mathrm{Tr}\left[(\bm{h} + \bm{\mathcal{F}}_\theta) \bm{\rho}_\theta - \bm{\Delta}_\theta \bar{\bm{\kappa}}^\star_\theta\right],
\end{split}
\end{equation}
where $\bm{\mathcal{F}}_\theta$ is a generalized Fock operator and $\bm{\Delta}_\theta$ and $\bar{\bm{\Delta}}_\theta$ are generalizations of the pairing matrix of HFB and CPMFT:
\begin{subequations}
\begin{align}
\mathcal{F}_{ik}(\theta) &= h_{ik} + \sum \langle ij \| kl \rangle \rho_{lj}(\theta)
\\
                         &= h_{ik} + G_{ik}(\theta)
\\
\Delta_{ij}(\theta) &= \frac{1}{2} \sum \langle ij \| kl \rangle \kappa_{kl}(\theta)
\\
\bar{\Delta}_{ij}(\theta) &= \frac{1}{2} \sum \langle ij \| kl \rangle \bar{\kappa}_{kl}(\theta).
\end{align}
\end{subequations}

The overlap matrix elements and the transition density matrices can be explicitly written in terms of the density matrix $\bm{\rho}$ and the anomalous density matrix $\bm{\kappa}$ of the underlying HFB state $|\Phi\rangle$.\cite{sheikh2000} In particular, the overlap matrix elements are written as
\begin{equation}
\langle \Phi | \hat{R}(\theta) | \Phi \rangle = \pm \frac{\mathrm{det} \, \mathbf{R}_\theta}{\sqrt{\mathrm{det} \, \bm{\rho}} \sqrt{\mathrm{det} \, \mathbf{C}_\theta}},
\end{equation}
where the matrix $\mathbf{C}_\theta$ is constructed from $\bm{\rho}$ and $\bm{\kappa}$ according to
\begin{equation}
\mathbf{C}^{-1}_\theta = \bm{\rho} \mathbf{R}_\theta \bm{\rho} \mathbf{R}^\dagger_\theta - \bm{\kappa} \mathbf{R}^\star_\theta \bm{\kappa}^\star \mathbf{R}^\dagger_\theta.
\end{equation}
The transition density matrices are in turn expressed as
\begin{subequations}
\begin{align}
\bm{\rho}_\theta &= \mathbf{R}_\theta \bm{\rho} \mathbf{R}^\dagger_\theta \mathbf{C}_\theta \bm{\rho},
\\
\bm{\kappa}_\theta &= \mathbf{R}_\theta \bm{\rho} \mathbf{R}^\dagger_\theta \mathbf{C}_\theta \bm{\kappa},
\\
\bar{\bm{\kappa}}^\star_\theta &= \mathbf{R}^\star_\theta \bm{\kappa}^\star \mathbf{R}^\dagger_\theta \mathbf{C}_\theta \bm{\rho}.
\end{align}
\end{subequations}
Since the matrix elements are all functionals of $\bm{\rho}$ and $\bm{\kappa}$, so too is the projected HFB energy itself:
\begin{equation}
E_j = E_j[\bm{\rho},\bm{\kappa}].
\end{equation}
We remind the reader that the index $j$ labels the quantum numbers which have been projectively restored.

\subsection{Variational Equations}
The importance of the foregoing result cannot be overemphasized. The fact that all matrix elements can be expressed in terms of the density and pairing matrices of the underlying HFB state implies that one can minimize the energy directly with respect to $\bm{\rho}$ and $\bm{\kappa}$, obtaining an effective mean-field Hamiltonian analogous to the Fock operator or the quasiparticle Hamiltonian of HFB.  Specifically, one can optimize the functional
\begin{equation}
\mathcal{L}[\bm{\rho},\bm{\kappa}] = E_j[\bm{\rho},\bm{\kappa}] - \mathrm{Tr} \left[\bm{\Lambda}\left(\bm{\mathcal{R}}^2 - \bm{\mathcal{R}}\right)\right]
\end{equation}
where $\bm{\Lambda}$ is a matrix of Lagrange multipliers used to constrain the generalized density matrix $\bm{\mathcal{R}}$, expressed in the Valatin form as
\begin{equation}
\bm{\mathcal{R}} = 
\begin{pmatrix}
 \bm{\rho} & \bm{\kappa} \\ -\bm{\kappa}^\star & \bm{1} - \bm{\rho}^\star
\end{pmatrix},
\end{equation}
to remain idempotent. The idempotency of the generalized density matrix is equivalent to the requirement that the state remains a quasiparticle Slater determinant.  Making the functional stationary with respect to variations in $\bm{\mathcal{R}}$ leads to the condition
\begin{equation}
[\bm{\mathcal{H}}_j,\bm{\mathcal{R}}] = 0,
\label{phfbeq}
\end{equation}
which is simply the Brillouin condition for HFB.  Here, the effective Hamiltonian matrix $\bm{\mathcal{H}}_j$ is given by
\begin{equation}
\bm{\mathcal{H}}_j = \frac{\delta E_j}{\delta \bm{\mathcal{R}}}.
\end{equation}
Expressions for the matrix elements of $\bm{\mathcal{H}}_j$ are provided in appendix A.  

Equation \ref{phfbeq} thus constitutes the form of the projected HFB equations expressed in matrix form. Solving the PHFB equations is thus conceptually simple. Given an initial guess of the quasiparticle density matrix $\bm{\mathcal{R}}$, one constructs the effective Hamiltonian matrix $\bm{\mathcal{H}}_j$, diagonalizes it, and constructs an updated $\bm{\mathcal{R}}$ using the eigenvectors of $\bm{\mathcal{H}}_j$. Convergence is achieved once Eqn. \ref{phfbeq} is satisfied up to a previously determined tolerance.

There are some subtle differences between the projected and the regular HFB variational problems that merit a few words. In the regular HFB equations, one must introduce the chemical potential $\mu$ forcing the HFB determinant to contain the correct particle number on average.  This chemical potential is not required in projected HFB, but we include it nonetheless as it often improves convergence of the projected HFB equations.  The chemical potential, to be clear, is applied to force the quasiparticle determinant to contain the correct number of electrons, which improves convergence of the projected HFB equations but does not change the final result.

There is also the key question of how one should occupy the quasiparticle orbitals after diagonalization of the projected HFB Hamiltonian. This has been discussed in the literature in the case of HFB.\cite{bertsch2009}  We emphasize that occupying the lowest-energy orbitals in the projected HFB Hamiltonian spectrum need not lead to convergence, or to the lowest energy solution even when convergence is achieved. Using the fact that, if the initial guess is good enough, then Eqn. \ref{phfbeq} is approximately satisfied at every cycle, we diagonalize the modified Hamiltonian matrix
\begin{equation}
\tilde{\bm{\mathcal{H}}}_j = \bm{\mathcal{H}}_j + \lambda \, \bm{\mathcal{R}},
\end{equation}
with $\lambda < 0$. When the equations are converged, the spectrum of $\tilde{\bm{\mathcal{H}}}_j$ is identical to that of $\bm{\mathcal{H}}_j$ except that the occupied quasiparticle orbital energies have been shifted down by $\lambda$. This ``level shifting''\cite{natiello1984} allows us to occupy the desired orbitals following an aufbau principle.  At self-consistency, it is of course irrelevant whether we have forced $\bm{\mathcal{R}}$ to commute with $\bm{\mathcal{H}}$ or with $\tilde{\bm{\mathcal{H}}}$.

\subsection{Simplified Forms of the Variational Equations}
We have written the projected HFB equations, Eqn. \ref{phfbeq} in the spin-orbital framework, allowing for quite general symmetry breaking.  However, the HFB and PHFB equations, like the HF equations, have self-consistent symmetries.  In other words, if we choose an initial density matrix to contain certain symmetries, then the self-consistent density matrix will also contain those symmetries.  Thus, for example, if we were to supply projected HFB with an RHF initial guess, we would converge to an RHF solution as RHF is a special case of the more general projected HFB.  It is up to the user, therefore, to decide which symmetries, if any, are to be broken in the unprojected HFB state.  Once this choice has been made, the procedure for restoring these symmetries is fairly general.

We can impose self-consistent symmetries on the underlying HFB state to simplify the PHFB equations, and discuss two such forms which we have used in this paper.  Yamaki \textit{et al}.\cite{yamaki2004} have discussed other possible structures of simplified problems in the context of regular HFB.

In the restricted HFB (RHFB) equations with singlet pairing, we prepare our initial guess for the quasiparticle density matrix $\bm{\mathcal{R}}$ imposing the following conditions: $\bm{\rho}_{\alpha\alpha} = \bm{\rho}_{\beta\beta}$, $\bm{\rho}_{\alpha\beta} = \bm{\rho}_{\beta\alpha} = \bm{0}$, $\bm{\kappa}_{\alpha\alpha} = \bm{\kappa}_{\beta\beta} = \bm{0}$, $\bm{\kappa}_{\alpha\beta} = (\bm{\kappa}_{\alpha\beta})^\mathsf{T}$.  Note that this form is appropriate for closed shell systems.  One can then solve the projected HFB equations using only half the dimension with the simplified quasiparticle density matrix $\tilde{\bm{\mathcal{R}}}$, given by
\begin{equation}
\tilde{\bm{\mathcal{R}}} = 
\begin{pmatrix}
\bm{\rho}_{\alpha\alpha}  & \bm{\kappa}_{\alpha\beta} \\ \bm{\kappa}^\star_{\alpha\beta} & \bm{1} - \bm{\rho}^\star_{\alpha\alpha}
\end{pmatrix},
\end{equation}
and the simplified effective Hamiltonian
\begin{equation}
\tilde{\bm{\mathcal{H}}}_j = \frac{\delta E_j}{\delta \tilde{\bm{\mathcal{R}}}}.
\end{equation}
Generally the RHFB wave function does not break spatial symmetry, but it retains the flexibility to do so.

The unrestricted HFB (UHFB) equations allow for spin symmetry breaking, though the overall HFB state remains an eigenfunction of $\hat{S}_z$. Here, we prepare our initial guess for the quasiparticle density matrix assuming the following conditions: $\bm{\rho}_{\alpha\beta} = \bm{\rho}_{\beta\alpha} = \bm{0}$, $\bm{\kappa}_{\alpha\alpha} = \bm{\kappa}_{\beta\beta} = \bm{0}$.  This leads to the following structure for the simplified quasiparticle density matrix:
\begin{equation}
\tilde{\bm{\mathcal{R}}} = 
\begin{pmatrix}
\bm{\rho}_{\alpha\alpha} & \bm{\kappa}_{\alpha\beta} \\ \bm{\kappa}_{\alpha\beta}^\dagger & \bm{1} - \bm{\rho}^\star_{\beta\beta}
\end{pmatrix}
\end{equation}
This allows for opposite spin ($m_S = 0$) triplet pairing\cite{TripletPairing} in the underlying quasiparticle determinant while excluding same spin ($m_S = \pm 1$) triplet pairing.  Breaking the $\hat{S}_z$ symmetry of the underlying HFB state introduces same-spin triplet pair correlations to the underlying quasiparticle determinant, and requires solving the PHFB equations in the general framework, which we denote as GHFB.

\section{Computational Details}
We have implemented the projected Hartree--Fock--Bogoliubov equations in the way described above both in an in-house code and as part of the {\small {GAUSSIAN}}\cite{gdv} suite of programs. We have validated our implementation of the particle-number projected equations by comparing results with an AGP code that we had available in our research group.\cite{staroverov2002}  We compared the energy, the natural orbital occupations and the geminal coefficients and reached quantitative agreement in all cases tested.

In this work, we only report tests on a few small singlet systems, using basis sets of minimal to double-$\zeta$ $+$ polarization quality.  Calculations on larger systems and with larger basis sets are possible, but we present here pilot calculations to illustrate the main features of the method.

\begin{table}[t]
\caption{Nomenclature defining our various projected and unprojected HFB states.
\label{Tab:Nomenclature}}
\begin{ruledtabular}
\begin{tabular}{lll}
Designation    &  Form of $\bm{\rho}$    &   Form of $\bm{\kappa}$
\\
\hline
RHFB
     &
$\begin{pmatrix}
\bm{\rho}_{\alpha\alpha}  &  \bm{0}   \\
\bm{0}     &  \bm{\rho}_{\alpha\alpha}
\end{pmatrix}$
     &
$\begin{pmatrix}
\bm{0}        &  \bm{\kappa}_{\alpha\beta}  \\
-\bm{\kappa}_{\alpha\beta}  &  \bm{0}
\end{pmatrix}$,
$\bm{\kappa}_{\alpha\beta} = \bm{\kappa}_{\alpha\beta}^\mathsf{T}$
     \\
UHFB
     &
$\begin{pmatrix}
\bm{\rho}_{\alpha\alpha} & \bm{0}    \\
\bm{0}                 & \bm{\rho}_{\beta\beta}
\end{pmatrix}$
     &
$\begin{pmatrix}
\bm{0}                    & \bm{\kappa}_{\alpha\beta}    \\
\bm{\kappa}_{\beta\alpha}   & \bm{0}
\end{pmatrix}$
     \\
GHFB
     &
$\begin{pmatrix}
\bm{\rho}_{\alpha\alpha} & \bm{\rho}_{\alpha\beta}    \\
\bm{\rho}_{\beta\alpha}  & \bm{\rho}_{\beta\beta}
\end{pmatrix}$
     &
$\begin{pmatrix}
\bm{\kappa}_{\alpha\alpha} & \bm{\kappa}_{\alpha\beta}    \\
\bm{\kappa}_{\beta\alpha}  & \bm{\kappa}_{\beta\beta}
\end{pmatrix}$
     \\
\hline
\\
Designation    &  \multicolumn{2}{l}{Symmetry Restored}
\\
\hline
N              &  \multicolumn{2}{l}{Particle Number}
\\
S              &  \multicolumn{2}{l}{Spin}
\\
K              &  \multicolumn{2}{l}{Complex Conjugation}
\\
$C_i$          &  \multicolumn{2}{l}{Inversion Point Group Symmetry}
\end{tabular}
\end{ruledtabular}
\end{table}

We have a wide variety of possible states, depending on which symmetries we wish to break in the underlying quasiparticle determinant and which broken symmetries we wish to restore by projection.  We summarize our nomenclature in Table \ref{Tab:Nomenclature}.  Note that we have not exhausted all  possible structures of the quasiparticle determinant $|\Phi\rangle$ and, therefore, the density matrices $\bm{\rho}$ and $\bm{\kappa}$.\cite{yamaki2004}  The versions we have used closely resemble the forms of Hartree-Fock determinants most commonly employed.

Let us pause to discuss the hierarchy of wave functions that fall into our classification.  Starting from a broken symmetry HFB determinant, number projection results in a single AGP wave function, with broken symmetry if the quasiparticle determinant is of UHFB or GHFB character.  Adding spin projection yields a linear combination of AGP wave functions such that the overall result is a spin eigenfunction, and restoring discrete symmetries then yields a linear combination of these wave functions.  It is not yet entirely clear how one could \textit{a priori} determine which symmetries must be broken and restored in any particular calculation or which of these symmetries is more important in one molecule than another. Future calculations will surely clarify this and other aspects of the methodology that we present here.

We prepare the initial guess to our restricted number-projected (NRHFB) calculations by diagonalizing the core Hamiltonian (neglecting the electron-electron repulsion) or by solving restricted HF equations. We use the orbital energies obtained to thermalize our initial guess of occupations according to a Fermi-Dirac distribution. We then build the singlet-pairing $\bm{\kappa}_{\alpha\beta}$ matrix using $\bm{\kappa}_{\alpha\beta} = \bm{\rho}_{\alpha\alpha} - \bm{\rho}_{\alpha\alpha}^2$.

For our spin symmetry broken calculations, our current approach is simple: we converge the restricted equations, and then mix the highest-occupied and the lowest-unoccupied quasiparticle eigenvectors using some predefined angle. Constructing the quasiparticle density matrix from such a set of eigenvectors leads to a spin-symmetry broken guess of $\bm{\mathcal{R}}$, that we then iterate to convergence.  The quality of this initial guess is rather poor, and improved initial guesses should help alleviate the convergence difficulties described below.

The number of grid points required in the discretization of the particle-number and spin projection integrations is relatively low. When using the simplified form of the equations appropriate for particle-number projection,\cite{sheikh2000} we have observed that convergence of the energy to 1 nano-Hartree is achieved with $7$ or $9$ grid points. Use of the more general equations presented in this paper seems to require a larger grid even for particle number projection, and we have used $15$ points for most of our calculations. Discretization over the Euler angles in spin projection leads to good convergence of our calculations when using about $10$ grid points per angle.  We generally integrate with equally spaced grid points using the trapezoid rule but use Gauss-Legendre quadrature for the integration over $\beta$ in the triaxial spin projection.  We discuss the accuracy of the integration grid in Section \ref{Sec:Grid}.  We emphasize that the calculations at each grid point are independent, and trivially parallel.

Finally, we should point out that the self-consistent equations of PHFB can be rather difficult to converge and we have employed the direct inversion of the iterative subspace (DIIS) algorithm\cite{pulay1982,GusDIIS} in addition to the level shifting described previously.

\section{Results and Discussion}
\subsection{Number Projection and AGP}
The simplest case of projected quasiparticle theory is number projection of a restricted HFB determinant, which we refer to as NRHFB and which is identical to AGP with singlet geminals.  As mentioned earlier, we have used this fact to test the correctness of our NRHFB implemention.  More interestingly, however, we can use our NRHFB code to generate AGP wave functions at mean field cost.  To the best of our knowledge, the previously most efficient formulations of AGP scale as $\mathcal{O}(M^5)$.\cite{weiner2002,mazziotti2001}

We note that the NRHFB/AGP wave function dwells strictly in the seniority 0 sector of Hilbert space and is a linear combination of each and every seniority 0 determinant with, however, restrictions relating the coefficients.  In other words, every determinant in the NRHFB is a closed shell, and each closed shell determinant is assigned a coefficient.

\begin{figure}[t]
\includegraphics[width=0.45\textwidth]{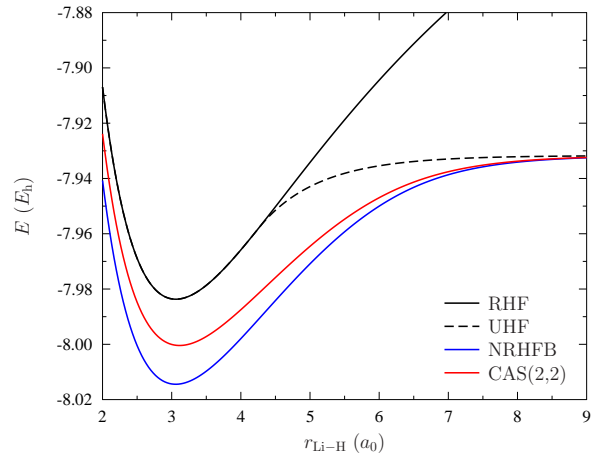}
\caption{Dissociation of LiH in the cc-pVDZ basis.  NRHFB is equivalent to AGP, and offers sizable improvement over RHF and UHF.
\label{Fig:LiH}}
\end{figure}

To illustrate the efficacy of NRHFB, we show the dissociation of LiH in Fig. \ref{Fig:LiH}.  As it is well known, AGP is exact for one electron pair, so LiH is perhaps the simplest dissociation case for which the method is not FCI.  As always, the RHF wave function is unable to dissociate to open-shell fragments, and the broken symmetry, spin-contaminated UHF solution splits off from it and dissociates to two UHF atoms.  Meanwhile, NRHFB offers significant dynamic correlation at equilibrium and dissociates cleanly to two ROHF atoms.  The CASSCF(2,2) is the minimal CAS needed to dissociate the molecule properly to ROHF fragments, and is included for comparison.  In this particular case, NRHFB is below the minimal CAS, although as we shall see this is not always so.

\subsection{Unrestricted HFB and AGP}
Perhaps the simplest extension to NRHFB is to allow the breaking of spin symmetry to give us NUHFB.  This is equivalent to an AGP with unrestricted (and potentially broken symmetry) orbitals.  As pointed out by Weiner \textit {et al}.\cite{weiner1984}, this is likely to be beneficial for the description of molecular dissociation, but to the best of our knowledge this is the first time that variationally optimized AGP with broken symmetry orbitals has been reported.  We should also add that simply by breaking spin symmetry, we allow the NUHFB wave function to permeate all of Hilbert space.  In other words, it spreads beyond the seniority 0 sector and gives us contributions from, in principle, every Slater determinant with the appropriate numbers of $\alpha$-spin and $\beta$-spin electrons.

\begin{figure}[t]
\includegraphics[width=0.45\textwidth]{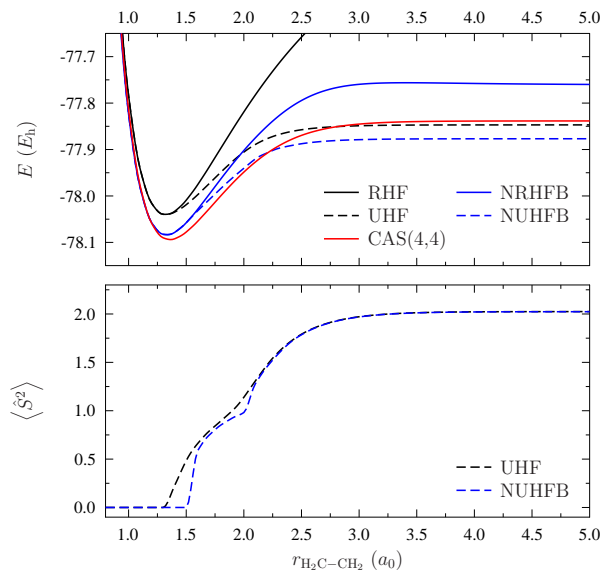}
\caption{Top Panel: Dissociation of C$_2$H$_4$ to two CH$_2$ fragments, in the cc-pVDZ basis.  Symmetry-broken AGP (\textit{i.e.} NUHFB) offers significant variational improvements.
Bottom Panel: Spin contamination in UHF and NUHFB as a function of C-C bond length.
\label{Fig:C2H4Dis}}
\end{figure}

In Fig. \ref{Fig:C2H4Dis} we show the dissociation of C$_2$H$_4$ to two triplet CH$_2$ fragments.  Because we are breaking a double bond, one can expect correlation effects to be very important.  One consequence is that the UHF curve separates from the RHF curve near equilibrium.  A second consequence is that NRHFB ends up between the RHF and UHF limits (though much closer to the latter).  In order to get a reasonable description of the dissociation, we are forced to break spin symmetry to give us NUHFB, which goes to a dissociation limit below two ROHF fragments.  Interestingly, the NUHFB curve is nearly parallel to the UHF curve, except near equilibrium where it picks up a little extra correlation.  We suspect that the dissociation limit amounts to one UHF fragment and one AGP fragment, though we cannot confirm this as we do not at present have the capability to do open-shell AGP calculations.  The minimal CAS is also shown; in this case it provides a slight variational improvement at equilibrium and dissociates to two ROHF fragments.  We can measure the spin contamination introduced by the broken symmetry NUHFB wave function, and Fig. \ref{Fig:C2H4Dis} also shows the expectation value of $\hat{S}^2$using the formulas provided in Appendix B.  It is interesting to note that for large bond lengths the NUHFB and UHF wave functions contain identical amounts of spin contamination.  The bumps in the UHF and NUHFB curves are manifestations of spatial symmetries being broken as the bond is stretched.  We note that the physical dissociation of C$_2$H$_4$ to triplet CH$_2$ fragments is accompanied by substantial changes in the geometry of the CH$_2$ fragments which we have here ignored but which are essential if one is to accurately reproduce experimental results.\cite{davidson1985}

\begin{figure}[t]
\includegraphics[width=0.45\textwidth]{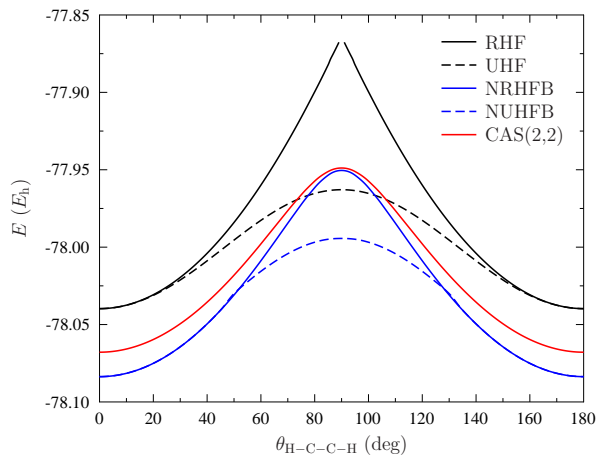}
\caption{Torsion of C$_2$H$_4$ about the C-C double bond, in the cc-pVDZ basis.  Breaking spatial and spin symmetry offers significant amounts of static correlation at both the HF and projected HFB levels.
\label{Fig:C2H4Rot}}
\end{figure}

A second example where spin symmetry breaking plays an important role is in the torsional barrier of C$_2$H$_4$, shown in Fig. \ref{Fig:C2H4Rot}.  This is a familiar multireference problem, where the correct curve requires two determinants at $\theta = 90^\circ$.  The RHF curve we have shown predicts a very large barrier, which is greatly reduced by UHF.  Similarly, the barrier in NRHFB is predicted to be rather large, but much smaller with NUHFB which again closely parallels the UHF curve.  The minimal CAS predicts the barrier to be somewhere between that predicted by NRHFB and NUHFB.  Note that we have ignored non-adiabatic effects which should be relevant here.

In order to provide a more formal description of what we mean by breaking spatial and spin symmetry in the context of HFB, we return to the form of the quasiparticle determinant, as written in Eqn. \ref{Eqn:AGP}.  We recall that the orbitals to be paired, represented by indices $k$ and $\bar{k}$, are usually the $\alpha$ and $\beta$ spin orbitals corresponding to the same spatial orbital.  Breaking spin symmetry means relaxing this constraint.  In particular, when we consider general spin orbitals\cite{GHF} to construct the quasiparticle vacuum, the spin label is lost as the orbitals have contributions from both $\alpha$ and $\beta$ character. In such a case, it is still true that $n_k = n_{\bar{k}}$, and those correspond to the orbitals paired, but it is no longer true that one of the paired orbitals is $\alpha$ and the other $\beta$. In breaking the spin symmetry of the wave function we have broken the time-reversal character of the Cooper pairs.

\subsection{Restoring Spin Symmetry}
Having broken spin symmetry, the next step is to restore it, yielding what we call SNUHFB and SNGHFB.  In both cases the projected HFB wave function is an eigenfunction of $\hat{S}^2$ and of $\hat{S}_z$.  The distinction between them is that the unprojected UHFB wave function is an eigenfunction of $\hat{S}_z$, while the unprojected GHFB wave function is not.  We emphasize that the RHFB wave function is an eigenfunction of both $\hat{S}^2$ and of $\hat{S}_z$; consequently, spin projection on the RHFB wave function merely returns the RHFB wave function.

\begin{figure}[t]
\includegraphics[width=0.45\textwidth]{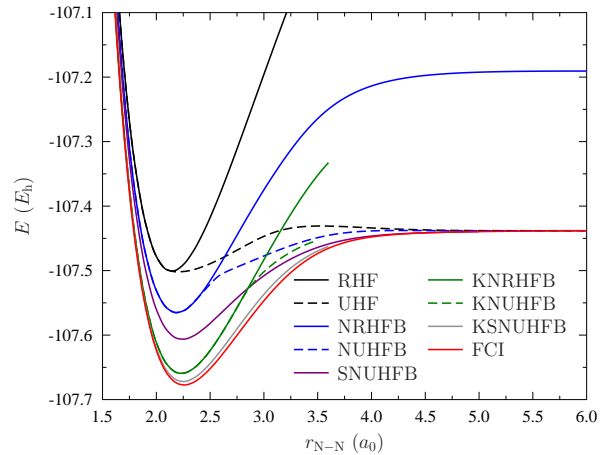}
\caption{Dissociation of N$_2$ in the STO-3G basis.  Restoring the broken spin symmetry provides significant correlation and eliminates the instability in the NRHFB wave function.
\label{Fig:N2}}
\end{figure}

Figure \ref{Fig:N2} shows the dissociation of N$_2$.  As is well known, the RHF curve is disastrous, while UHF strongly underbinds and predicts a barrier to the formation of the N-N bond.  As with the dissociation of C$_2$H$_4$, NRHFB goes to a dissociation limit between the RHF and UHF limits while offering some improvement at equilibrium.  In this case, NUHFB follows NRHFB near equilibrium but goes to the UHF limit at dissociation; that NUHFB does not go below the UHF limit at dissociation is simply because we are working in a minimal basis.  Restoring the broken spin symmetry in SNUHFB offers substantial improvements all across the dissociation curve while still dissociating to two ROHF atoms.

\subsection{Restoring Complex Conjugation Symmetry}
Figure \ref{Fig:N2} also shows the dissociation curves predicted by KNRHFB and KNUHFB for the nitrogen molecule. Complex conjugation restoration recovers a very significant fraction of correlation near equilibrium.  KNRHFB does not dessociate to the correct limit, a feature that KNUHFB solves by breaking spin symmetry.  Adding complex conjugation on top of spin projection (KSNUHFB) yields a curve that stays very close to full CI for all bond lengths.

\begin{figure}[t]
\includegraphics[width=0.45\textwidth]{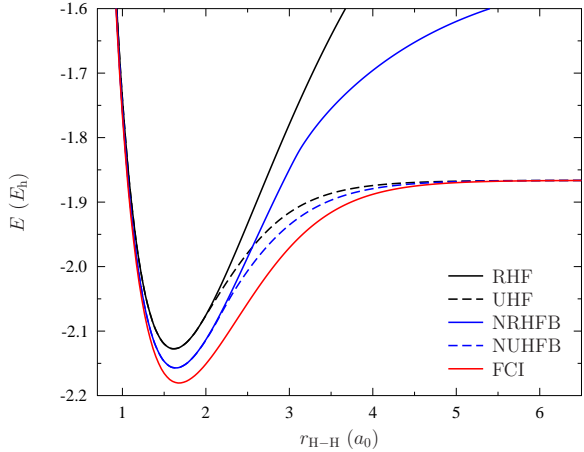}
\caption{Dissociation of equally spaced linear H$_4$ in the STO-3G basis.  Only by breaking symmetry can projected HFB reach the correct dissociation limit.
\label{Fig:H4}}
\end{figure}

A particularly interesting test case is provided by an equally spaced linear chain of hydrogen atoms.  We here limit ourselves to H$_4$ in a minimal basis.  As can be seen in Fig. \ref{Fig:H4}, RHF dissociates incorrectly while UHF yields four hydrogen atoms.  As we have come to expect, NRHFB (even with broken spatial symmetry) gives a dissociation limit between RHF and UHF, while NUHFB goes to the UHF dissociation limit. 

\begin{figure}[t]
\includegraphics[width=0.45\textwidth]{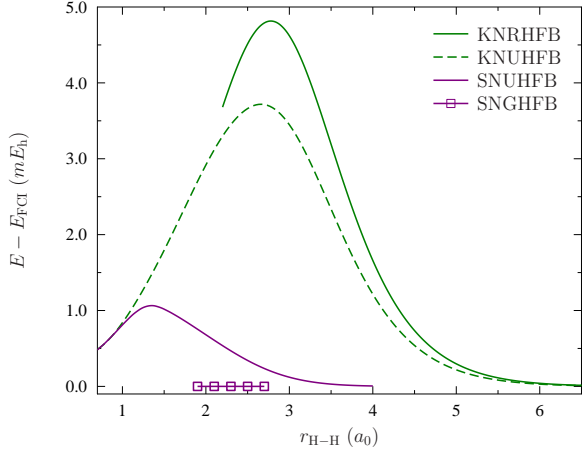}
\caption{Comparison of KNHFB and SNHFB to full CI for the dissociation of equally spaced linear H$_4$.  Restoring these symmetries leads to results almost equal to full CI.  In particular, SNGHFB appears to be energetically identical to full CI for the points at which we successfully converged the equations.
\label{Fig:H4b}}
\end{figure}

Restoring additional symmetries leads to results nearly identical to full CI, and Fig. \ref{Fig:H4b} therefore shows the deviation from full CI.  We achieve significant improvement over NUHFB by restoring collinear spin symmetry.  Restoring noncollinear spin symmetry in NGHFB appears to yield results agreeing with full CI to better than the nano-Hartree level, though we have experienced great difficulties in converging the equations and have been unable to generate the entire potential energy curve.  We speculate that the accuracy of the SNGHFB is tied in some way to the fact that we have only two symmetry-unique hydrogen atoms (and thus only two symmetry-unique electrons) and are working in a minimal basis.

Rather than restoring spin symmetry, we can choose to restore complex conjugation symmetry, and in so doing achieve the correct dissociation limit with restricted orbitals.

\subsection{Restoring Spatial Symmetry}
Throughout our discussion, the reader may have noticed that our projected HFB wave functions generally dissociate to ROHF fragments.  Of course projected HFB calculations on the isolated fragments would yield some correlation.  In other words, projected HFB is not size consistent since it does not dissociate to projected HFB fragments.  In the special case of NRHFB = AGP, this is well known.\cite{staroverov2002}  The lack of size consistency is not fatal for molecular applications, but is of course a serious problem for applications to periodic systems, where the correlation energy per unit cell would vanish as the size of the unit cell is increased.


\begin{figure}[t]
\includegraphics[width=0.45\textwidth]{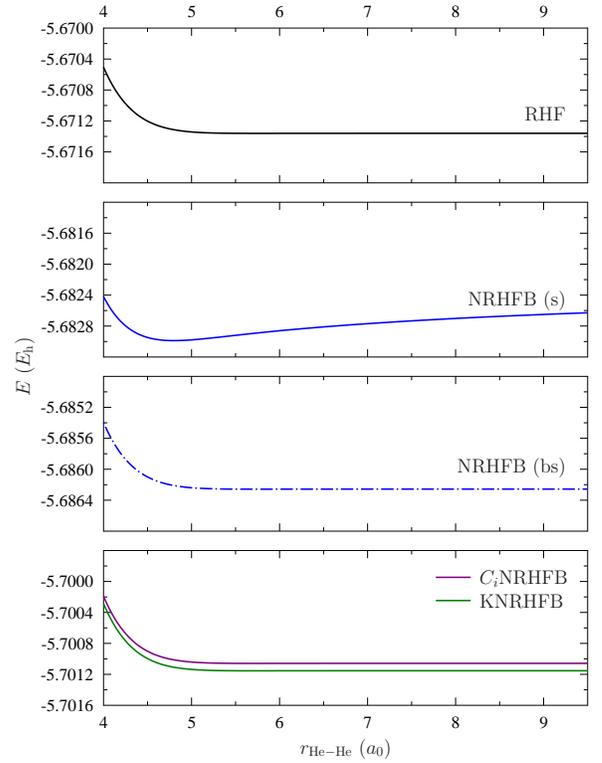}
\caption{Dissociation of He$_2$ in the 3-21G basis.  We have broken the energy axis several times as the solutions are well-separated in energy and the binding is minimal.  The full CI results exactly match the KNRHFB curve shown in the figure.  The labels "s" and "bs" on the NRHFB indicate whether spatial symmetry is preserved ("s") or broken ("bs") in the underlying HFB determinant.
\label{Fig:He2}}
\end{figure}

We consider the dissociation of He$_2$ in Fig. \ref{Fig:He2}, using the 3-21G basis.  The RHF curve is strictly repulsive, while NRHFB has noticeable binding strictly because it goes to the wrong dissociation limit.  This binding is greatly exaggerated, as the experimental binding energy is on the order of 30 microHartrees.  By breaking spatial symmetry, we obtain a curve that closely parallels the RHF curve; in fact, the dissociation limit corresponds to one RHF atom and one AGP atom.  This suggests that spatial symmetry restoration (here equivalent to restoring inversion symmetry due to the small basis being used) has something to offer.  In fact, when we restore spatial symmetry in $C_i$-NRHFB, we obtain a curve which is only about 0.1 milliHartree above full CI.  Curiously, restoring instead complex conjugation symmetry in KNRHFB gives results which are numerically identical to the FCI.

Comparing NRHFB to $C_i$NRHFB, we note that one recovers all but $\sim$ 0.1 milliHartree of the $\sim$ 15 milliHartrees error in the dissociation limit of the broken symmetry NRHFB, reducing the error by more than 99\%.  We suspect that by breaking and restoring the remaining possible symmetries we may be able to achieve FCI results.

\subsection{Accounting for Static Correlation}
In previous plots, we showed that the projected HFB wave function can describe \textit{left-right} correlation, a form of strong correlation that occurs in molecular dissociation to fragments. This can be accomplished by breaking the spatial and spin symmetry on top of the number projected wave function (as in the case of the NUHFB dissociation of N$_2$), or by breaking and restoring other symmetries (as in the case of the KNRHFB dissociation of H$_4$).  

Another type of strong correlation is referred to as \textit{angular},\cite{scuseria2009} and is related to the near degeneracies of orbitals with the same principal quantum number in atomic systems. For instance, the $2s$ and the $2p$ orbitals of beryllium are close in energy, and thus a multireference wave function is needed to give a correct description of the electronic structure of this atom.

\begin{figure}[t]
\includegraphics[width=0.45\textwidth]{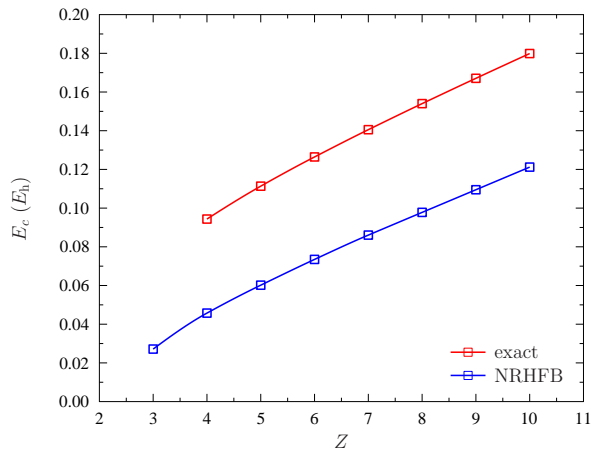}
\caption{Correlation energies in the four-electron series.  We compare the NRHFB results in the aug-cc-pVTZ basis to the exact results of Ref. \onlinecite{chakravorty1993}.  Note that we correctly recover the linear behavior of the correlation energy as a function of nuclear charge $Z$.
\label{Fig:4e}}
\end{figure}

We show in Fig. \ref{Fig:4e} the correlation energy predicted by the NRHFB method on the series of atomic species having 4 electrons.   We compare to the exact correlation energies at the non-relativistic, Born-Oppenheimer level of theory provided by Chakravorty \textit{et al}.\cite{chakravorty1993}.  We have used the aug-cc-pVTZ basis for NRHFB calculations; this basis is large enough to provide qualitatively correct behavior, but still departs from the complete basis set limit.  The leading contribution to the correlation energy is known to scale linearly with $Z$ for large enough $Z$,\cite{Linderberg1960} and NRHFB reproduces this behavior.

We note that if the core occupations were frozen (both in FCI and in NRHFB), then NRHFB would agree exactly with FCI because what remains is a two-electron system and AGP (or NRHFB) is exact for any two-electron singlet.

\subsection{Accuracy of Projections
\label{Sec:Grid}}
In order to assess the number of grid points required in the particle number projection, we have performed calculations on rings of hydrogen atoms of increasing size. In our results, we have converged NRHFB states with 9 grid points in the integration from 0 to $\pi$ using the trapezoidal rule and the equations derived in Ref. \onlinecite{sheikh2000}. (Since the number of particles is even, the integrand is symmetric about $\pi$.) We then evaluated errors in the energy and in the number of particles with grids of various sizes using the converged density matrix.

Our results are shown in Table \ref{tb:grid}. We note that the trapezoidal rule becomes exact when the number of grid points used satisfies $M = \mathrm{max} (\frac{1}{2} N, \Omega - \frac{1}{2} N) +1$, where $N$ is the total number of particles and $\Omega$ is the degeneracy of the space.\cite{hara1979} Thus, the trapezoidal grid becomes exact for H$_8$ in minimal basis with 5 grid points.

The numbers shown in Table \ref{tb:grid} suggest that the size of the grid required to achieve certain accuracy in the energy or in the variance of the number of particles scales less than linearly with the number of particles. We expect the same weak dependence for other projection operators. Presumably, the quality of the grid required for spin projection depends on the magnitude of the spin contamination present in the broken symmetry HFB state.

\begin{table*}
\caption{Errors associated with the discretization of the particle
number projection operator in NRHFB calculations of hydrogen atom
rings ($r_{\mathrm{H-H}} = 1.80$ bohr, STO-3G basis). The
reference energy is taken to be that with $N_{\mathrm{grid}} =
9$. \label{tb:grid}}
\begin{ruledtabular}
\begin{tabular}{lccccc}
Property  &  $N_{\mathrm{grid}}$  &  H$_8$  &  H$_{16}$  &  H$_{32}$  &  H$_{64}$  \\
\hline
$\mathrm{abs} \left( \left\langle N \right\rangle - N \right)$
          & 2
          & $7.53 \times 10^{-3}$
          & $3.05 \times 10^{-3}$ 
          & $1.26 \times 10^{-3}$
          & $5.71 \times 10^{-4}$ 
\\
          & 4
          & $4.88 \times 10^{-6}$
          & $1.46 \times 10^{-5}$
          & $1.43 \times 10^{-5}$
          & $9.06 \times 10^{-6}$ 
\\
          & 6
          & $< 10^{-10}$
          & $1.41 \times 10^{-9}$
          & $6.95 \times 10^{-9}$
          & $9.73 \times 10^{-9}$
\\
          & 8 
          & $< 10^{-10}$
          & $< 10^{-10}$         
          & $< 10^{-10}$
          & $< 10^{-10}$
\\
\\
$\mathrm{abs} \left( \left\langle N^2 \right\rangle - \left\langle N \right\rangle^2 \right)$
          & 2
          & $2.21 \times 10^{-1}$
          & $2.62 \times 10^{-1}$
          & $2.87 \times 10^{-1}$
          & $2.87 \times 10^{-1}$
\\
          & 4
          & $7.03 \times 10^{-5}$
          & $3.55 \times 10^{-4}$
          & $7.99 \times 10^{-4}$
          & $1.17 \times 10^{-3}$
\\
          & 6 
          & $< 10^{-10}$
          & $2.41 \times 10^{-8}$
          & $2.20 \times 10^{-7}$
          & $6.90 \times 10^{-7}$
\\
          & 8
          & $< 10^{-10}$
          & $< 10^{-10}$
          & $< 10^{-10}$
          & $1.15 \times 10^{-10}$
\\
\\
$\mathrm{abs} \left( \left\langle H \right\rangle - E_{\mathrm{ref}} \right)$
          & 2
          & $5.92 \times 10^{-2}$
          & $5.46 \times 10^{-2}$
          & $4.78 \times 10^{-2}$
          & $4.13 \times 10^{-2}$
\\
          & 4
          & $1.54 \times 10^{-5}$
          & $5.91 \times 10^{-5}$
          & $9.69 \times 10^{-5}$
          & $1.10 \times 10^{-4}$
\\
          & 6
          & $< 10^{-10}$
          & $3.53 \times 10^{-9}$
          & $2.32 \times 10^{-8}$
          & $5.33 \times 10^{-8}$
\\
          & 8
          & $< 10^{-10}$
          & $< 10^{-10}$
          & $< 10^{-10}$
          & $< 10^{-10}$
\end{tabular}
\end{ruledtabular}
\end{table*}

\section{Concluding Remarks}
Symmetry breaking plays a central role in our current understanding of classical and quantum systems.\cite{strocchi2008} In finite electronic structure systems, spontaneous symmetry breaking in the presence of exact or near degeneracies is an artifact and quantum fluctuations (strong correlations) remove them.  In the presence of degeneracies, strong correlation therefore cannot be neglected if one wants to obtain a correct qualitative picture.\cite{yannouleas2007} One should remark that Schr\"odinger's equation is linear whereas the mean field equations of HF and HFB theories are cubic in the orbitals, and have more solutions than the physical ones.  Spontaneous symmetry breaking in mean-field theories flags emerging behavior, the appearance of phenomena (due to degeneracies) that requires a description beyond the single determinant picture.  By breaking symmetry, mean-field theories lift the degeneracy (\textit{e.g.} the HOMO-LUMO gap opens going from RHF to UHF in H$_2$ near dissociation) and can sometimes give qualitative descriptions of these phenomena at the cost of good quantum numbers.  In this sense, mean-field theories predict their own failure and signal the need for a more comprehensive treatment. Projection after variation is not the answer because orbital relaxation effects are very important. With PAV, the unphysical mean-field behavior is frequently enhanced rather than eliminated.\cite{schlegel1986,GHF} On the other hand, the results in this paper seem to indicate that a comprehensive variation-after-projection treatment of all molecular symmetries is capable of accounting for molecular static correlation in a black-box manner that yields smooth dissociation curves. The fact that this can be achieved with mean field computational cost is truly remarkable and unprecedented. On the other hand, extended systems behave differently from finite systems and it is usually argued that spontaneous symmetry breaking there has the physically meaningful interpretation associated with true phase transitions.\cite{anderson1984,blaizot1985}

The theory presented in this paper covers important unexplored aspects of electronic structure theory. The first one is the underlying coherent state representation where the generator coordinate method projects out variational states of the correct symmetry.\cite{ring1980,blaizot1985} The manifold of states from which this projection is performed is non-orthogonal and overcomplete. Our projected quasiparticle states are multireference wave functions. A second salient aspect of our work is the connection with geminal theories.\cite{coleman2000}  We have extended these geminal wave functions to variationally include unrestricted and general spin orbitals. We have also calculated wave functions that are linear combinations of these general-orbital AGPs.

Compared to CPMFT, our previous model for strong correlations, the current theory is N-representable and possesses a two-particle density matrix that is factorizable over the gauge grid (see Appendix C). As a matter of fact, all higher order reduced density matrices factorize in a similar way. This property will certainly be of interest to other workers who are building correlation models on top of multireference descriptions that do not possess this very useful factorization, since the factorization leads to one power reduction in computational scaling (typically from $\mathcal{O}(M^6)$ to $\mathcal{O}(M^5)$).

Regarding the separation of static and dynamic correlation that we have previously advocated in our CPMFT work,\cite{tsuchimochi2009,scuseria2009,tsuchimochi2010} one should note that the present theory is exact for any two-electron system. This, in our previous definition, includes both static and dynamic correlation. In the present geminal context, it seems more advantageous to describe correlations as originating from inter- and intra-pair interactions. It is evident that PHFB includes some but not all dynamical correlations although it seems to include all strong correlations for molecules.

One more time, recapitulating the main points of this work:
\begin{itemize}
\item First, we work with underlying unprojected \textit{quasiparticle} determinants which deliberately break symmetries of the exact wave function.  These symmetries are restored by projection to give projected quasiparticle wave functions which are obtained variationally.  We are following the procedure, that is, of variation-after-projection.  The projected quasiparticle wave functions are multireference in character.
\item Second, because the underlying unprojected wave function is simply a single determinant of quasiparticles, and the projected wave function is completely specified by the unprojected wave function, our projected wave function is specified by a regular density matrix $\bm{\rho}$ and an anomalous density matrix $\bm{\kappa}$.  The problem of variation after projection reduces to the problem of optimizing $\bm{\rho}$ and $\bm{\kappa}$ for the unprojected state, which can be accomplished at \textit{mean-field} computational cost.
\item The projection operators we are using are written as integrations over gauge angles in a generator coordinate approach.  We simply discretize these integrals to obtain numerically efficient projections, and fortunately the grids we need for each projection operator are not large.  We should point out, also, that while the computational cost is roughly equivalent to that of a mean field calculation at every grid point, the problem of gauge integration to do the projections is trivially parallel.
\end{itemize}

In summary, to the best of our knowledge the main new accomplishments presented here are:
\begin{itemize}
\item We have presented an efficient algorithm allowing for AGP calculations at mean field computational cost.
\item We have presented the first variational AGP calculations with broken symmetry orbitals.
\item We have reported the first full VAP calculations based on HFB for the combination of number, spin, and complex conjugation restoration.
\item We have reported the first discrete symmetry restoration (complex conjugation and point group) in quantum chemistry.
\item We have reported the first application of VAP with the full electronic Hamiltonian.
\end{itemize}

In closing, we would like to emphasize that the calculations presented in this work are just proof-of-principle benchmarks and only meant to demonstrate the compelling capability of the theory. Larger bases and chemically meaningful results will be presented in due time.

\section{Acknowledgments}
This work is supported by the National Science Foundation under CHE-0807194 and CHE-1110884, the Welch Foundation (C-0036), and Los Alamos National Labs (Subcontract 81277-001-10).

\appendix
\section{Equations for the Effective Hamiltonian}
Here we present the expressions for the effective Hamiltonian in the case of a general projection operator $\hat{P}$ associated with the rotation operator $\hat{R}(\theta)$.  We note that in their 2000 paper, Sheikh and Ring derived a simplified form of these equations applicable only to particle-number projection.

Let us begin by discussing the general form of the effective Hamiltonian.  Recall that the energy is given by
\begin{equation}
E = \frac{1}{2} \int \mathrm{d}\theta \, y(\theta) \, \mathrm{Tr}[(\bm{h} + \bm{\mathcal{F}}_\theta) \bm{\rho}_\theta - \bm{\Delta}_\theta \bar{\bm{\kappa}}^\star_\theta]
\end{equation}
which we will express equivalently as
\begin{equation}
E = \frac{1}{2} \int \mathrm{d}\theta \, y(\theta) \, \mathrm{Tr}[(2\bm{h} + \bm{G}_\theta) \bm{\rho}_\theta - \bm{\Delta}_\theta \bar{\bm{\kappa}}^\star_\theta].
\end{equation}
The effective Hamiltonian will take the general form
\begin{equation}
\bm{\mathcal{H}} = 
\begin{pmatrix}
\bm{F}^\rho + \bm{\Delta}^\rho                &   \bm{F}^\kappa + \bm{\Delta}^\kappa   \\
-(\bm{F}^\kappa + \bm{\Delta}^\kappa)^\star   &  -(\bm{F}^\rho + \bm{\Delta}^\rho)^\star
\end{pmatrix}
\end{equation}
where
\begin{subequations}
\begin{align}
F^\rho_{ij} &= \frac{\partial\hfill}{\partial \rho_{ji}} \frac{1}{2} \int \mathrm{d}\theta \, y(\theta) \, \mathrm{Tr}[(2\bm{h} + \bm{G}_\theta) \bm{\rho}_\theta],
\\
F^\kappa_{ij} &= -\frac{\partial\hfill}{\partial \kappa_{ji}^\star} \frac{1}{2} \int \mathrm{d}\theta \, y(\theta) \, \mathrm{Tr}[(2\bm{h} + \bm{G}_\theta) \bm{\rho}_\theta],
\\
\Delta^\rho_{ij} &= -\frac{\partial\hfill}{\partial \rho_{ji}} \frac{1}{2} \int \mathrm{d}\theta \, y(\theta) \, \mathrm{Tr}[\bm{\Delta}_\theta \bar{\bm{\kappa}}^\star_\theta],
\\
\Delta^\kappa_{ij} &= \frac{\partial\hfill}{\partial \kappa_{ji}^\star} \frac{1}{2} \int \mathrm{d}\theta \, y(\theta) \, \mathrm{Tr}[\bm{\Delta}_\theta \bar{\bm{\kappa}}^\star_\theta].
\end{align}
\end{subequations}
Dependence on $\bm{\rho}$ and $\bm{\kappa}$ lurks in $y(\theta)$ and in the transition density matrices $\bm{\rho}_\theta$, $\bm{\kappa}_\theta$, and $\bar{\bm{\kappa}}_\theta$ (and thus in $\bm{G}_\theta$, $\bm{\Delta}_\theta$, and $\bar{\bm{\Delta}}_\theta$).

Following Sheikh and Ring, we write the derivatives of the function $y(\theta)$ as
\begin{align}
\frac{\partial y(\theta)}{\partial \rho_{kl}} &= y(\theta) Y_{lk}^\rho(\theta),
\\
\frac{\partial y(\theta)}{\partial \kappa_{kl}^\star} &= y(\theta) Y_{lk}^{\kappa}(\theta) = y(\theta) [\tilde{Y}_{lk}(\theta) - \tilde{Y}_{kl}(\theta)].
\end{align}
We therefore obtain
\begin{subequations}
\begin{align}
F^\rho_{ij} &=  \int \mathrm{d}\theta \, y(\theta) \Big\{Y_{ij}^\rho \, \mathrm{Tr}[(\bm{h} + \frac{1}{2} \bm{G}_\theta) \bm{\rho}_\theta]
\\
           & \qquad +\sum \mathcal{F}_{lk}(\theta) \frac{\partial \rho_{kl}(\theta)}{\partial \rho_{ji}}\Big\}
\nonumber
\\
F^\kappa_{ij} &= - \int \mathrm{d}\theta \, y(\theta) \Big\{ Y_{ij}^\kappa \, \mathrm{Tr}[(\bm{h} + \frac{1}{2} \bm{G}_\theta) \bm{\rho}_\theta]
\\
           & \qquad + \sum \mathcal{F}_{lk}(\theta) \frac{\partial \rho_{kl}(\theta)}{\partial \kappa_{ji}^\star} \Big\}
\nonumber
\\
\Delta^\rho_{ij} &=  -\frac{1}{2} \int \mathrm{d}\theta \, y(\theta) \Big\{ Y_{ij}^\rho \, \mathrm{Tr}[\bm{\Delta}_\theta \bar{\bm{\kappa}}^\star_\theta]
\\
                &  \qquad + \sum \left[\Delta_{lk}(\theta) \frac{\partial \bar{\kappa}^\star_{kl}(\theta)}{\partial \rho_{ji}}
                                      + \bar{\Delta}^\star_{lk}(\theta) \frac{\partial \kappa_{kl}(\theta)}{\partial \rho_{ji}}\right]\Big\}
\nonumber
\\
\Delta^\kappa_{ij} &=  \frac{1}{2} \int \mathrm{d}\theta \, y(\theta) \Big\{ Y_{ij}^\kappa \, \mathrm{Tr}[\bm{\Delta}_\theta \bar{\bm{\kappa}}^\star_\theta]
\\
                & \qquad + \sum \left[\Delta_{lk}(\theta) \frac{\partial \bar{\kappa}^\star_{kl}(\theta)}{\partial \kappa^\star_{ji}}
                                     + \bar{\Delta}^\star_{lk}(\theta) \frac{\partial \kappa_{kl}(\theta)}{\partial \kappa^\star_{ji}}\right]\Big\}
\nonumber
\end{align}
\end{subequations}
We emphasize that $\bm{F}^\rho$, $\bm{F}^\kappa$, $\bm{\Delta}^\rho$, and $\bm{\Delta}^\kappa$ are generalizations of the usual Fock and pairing matrices of HFB and CPMFT.  The derivatives of the transition density matrices are involved and we merely provide the final expressions.  Note that $\bm{F}^\rho$, $\bm{\Delta}^\rho$, and $\mathbf{Y}^\rho$ are Hermitian, though this is not obvious from the matrix expressions below and we therefore Hermitize at the end.  Similarly, $\bm{F}^\kappa$, $\bm{\Delta}^{\kappa}$, and $\mathbf{Y}^\kappa$ are antisymmetric and we have explicitly antisymmetrized them in the course of taking the derivatives.

Our final results are that
\begin{align}
\mathbf{Y}^\rho_\theta
   &= \frac{1}{2} \left(\mathbf{R}_\theta \bm{\rho} \mathbf{R}_\theta^\dagger \mathbf{C}_\theta + \mathbf{R}_\theta^\dagger \mathbf{C}_\theta \bm{\rho} \mathbf{R}_\theta\right)
\\
   &- \frac{1}{2} \int \mathrm{d}\phi \, y(\phi) \left(\mathbf{R}_\phi \bm{\rho} \mathbf{R}_\phi^\dagger \mathbf{C}_\phi + \mathbf{R}_\phi^\dagger \mathbf{C}_\phi \bm{\rho} \mathbf{R}_\phi\right) 
\nonumber
\\
\tilde{\mathbf{Y}}_\theta &=  -\frac{1}{2} \mathbf{R}_\theta^\dagger \mathbf{C}_\theta \bm{\kappa} \mathbf{R}_\theta^\star + \frac{1}{2} \int\mathrm{d}\phi \, y(\phi)  \mathbf{R}_\phi^\dagger \mathbf{C}_\phi \bm{\kappa} \mathbf{R}_\phi^\star
\\
\mathbf{Y}^\kappa_\theta
  &= \tilde{\mathbf{Y}}_\theta -  \tilde{\mathbf{Y}}_\theta^\mathsf{T}
\end{align}

Using these, we have
\begin{subequations}
\begin{align}
\bm{F}^\rho &= \frac{1}{2} \Bigg\{ \int \mathrm{d}\theta \, y(\theta) \Big[ \bm{Y}^\rho \, \mathrm{Tr}[(\bm{h} +\frac{1}{2}  \bm{G}_\theta) \bm{\rho}_\theta]
\\
           & \qquad +  \mathbf{R}_\theta^\dagger \mathbf{C}_\theta \bm{\rho} \bm{\mathcal{F}}_\theta (\bm{1}-\bm{\rho}_\theta) \mathbf{R}_\theta
\nonumber
\\
          & \qquad + (\bm{1}-\bm{\rho}_\theta) \bm{\mathcal{F}}_\theta \mathbf{R}_\theta \bm{\rho}\mathbf{R}_\theta^\dagger \mathbf{C}_\theta \Big]\Bigg\} + \mathrm{h.c.}
\nonumber
\\
\bm{F}^\kappa  &= -\Bigg\{ \int \mathrm{d}\theta \, y(\theta) \Big[ \tilde{\bm{Y}} \, \mathrm{Tr}[(\bm{h} + \frac{1}{2} \bm{G}_\theta) \bm{\rho}_\theta]
\\
           & \qquad + \mathbf{R}_\theta^\dagger \mathbf{C}_\theta \bm{\rho} \bm{\mathcal{F}}_\theta \bm{\kappa}_\theta \mathbf{R}_\theta^\star\Big]\Bigg\} + (\cdots)^\mathsf{T}
\nonumber
\\
\bm{\Delta}^\rho &=  \frac{1}{4} \Bigg\{ \int \mathrm{d}\theta \, y(\theta) \Big[ -\bm{Y}^\rho \, \mathrm{Tr}[\bm{\Delta}_\theta \bar{\bm{\kappa}}^\star_\theta]
\\
                &  \qquad + \mathbf{R}_\theta^\dagger \mathbf{C}_\theta \bm{\rho} \bm{\Delta}_\theta \bar{\bm{\kappa}}_\theta^\star \mathbf{R}_\theta
\nonumber
\\
                & \qquad - (\bm{1}-\bm{\rho}_\theta) \bm{\Delta}_\theta \mathbf{R}_\theta^\star \bm{\kappa}^\star \mathbf{R}_\theta^\dagger \mathbf{C}_\theta
\nonumber
\\
                & \qquad - \mathbf{R}_\theta^\dagger \mathbf{C}_\theta \bm{\kappa} \bar{\bm{\Delta}}_\theta^\star (\bm{1}-\bm{\rho}_\theta) \mathbf{R}_\theta
\nonumber
\\
                & \qquad + \bm{\kappa}_\theta \bar{\bm{\Delta}}_\theta^\star \mathbf{R}_\theta \bm{\rho} \mathbf{R}_\theta^\dagger \mathbf{C}_\theta \Big] \Bigg\} + \mathrm{h.c.}
\nonumber
\\
\bm{\Delta}^\kappa &=  \frac{1}{2} \Bigg\{ \int \mathrm{d}\theta \, y(\theta) \Big[ \tilde{\bm{Y}} \, \mathrm{Tr}[\bm{\Delta}_\theta \bar{\bm{\kappa}}^\star_\theta]
\\
                  & \qquad +\mathbf{R}_\theta^\dagger \mathbf{C}_\theta \bm{\rho} \bm{\Delta}_\theta \bm{\rho}_\theta^\mathsf{T} \mathbf{R}_\theta^\star
\nonumber
\\
                  & \qquad + \mathbf{R}_\theta^\dagger \mathbf{C}_\theta \bm{\kappa} \bar{\bm{\Delta}}_\theta^\star \bm{\kappa}_\theta \mathbf{R}_\theta^\star\Big]\Bigg\} - (\cdots)^\mathsf{T}.
\nonumber
\end{align}
\end{subequations}

\section{Evaluation of $\langle \hat{S}^2 \rangle$}
In general, the expectation value of an operator $\hat{O}$ with our PQT wave 
functions is
\begin{equation}
\langle \hat{O} \rangle = \frac{\langle \Phi | \hat{P} \, \hat{O} \, \hat{P} | \Phi \rangle}{\langle \Phi | \hat{P} | \Phi \rangle}.
\end{equation}
Evaluating the numerator requires integration over the projection operator grid twice.  We have used this approach, for example, in evaluating density matrices (see below).  For the special case that the projection operator commutes with the operator $\hat{O}$, the expectation value simplifies to
\begin{equation}
\langle \hat{O} \rangle = \frac{\langle \Phi | \hat{O} \, \hat{P} | \Phi \rangle}{\langle \Phi | \hat{P} | \Phi \rangle}
\end{equation}
as we have used to evaluate the energy.

Because the spin projection operator takes the form $\exp(\mathrm{i} \alpha \hat{S}_z) \, \exp(\mathrm{i} \beta \hat{S}_y) \, \exp(\mathrm{i} \gamma \hat{S}_z)$, it does not commute with the individual $\hat{S}_i$, so evaluating their expectation values is most conveniently done in terms of the one-particle density matrix given below.  However, since $\hat{S}_y$ and $\hat{S}_z$ both commute with $\hat{S}^2$, evaluation of $\langle \hat{S}^2 \rangle$ is simpler.  We find that
\begin{align}
\langle \hat{S}^2 \rangle &= \int \mathrm{d}\theta \, y (\theta) \, 
 \Bigg\{\sum_i \bigg(\mathrm{Tr}\left[\mathbf{M}_i(\theta)\right]^2
\\
  & \qquad\qquad\qquad + \frac{1}{2} \mathrm{Tr}\left[\mathbf{M}^2_i(\theta) - \bar{\bm{\kappa}}_i^\star(\theta) \bm{\kappa}_i(\theta)\right]\bigg)
\nonumber
\\
  &\qquad \, +  \frac{3}{2} \, \mathrm{Tr}\left[\mathbf{P}(\theta) -  \mathbf{P}^2(\theta) - \bar{\bm{\kappa}}^\star_0(\theta) \, \bm{\kappa}_0(\theta)\right]\Bigg\}
\nonumber
\end{align}
where we have decomposed the spin blocks of the transition density matrices as
\begin{align}
\bm{\rho}(\theta)
  &\equiv  \begin{pmatrix} \bm{\rho}_{\alpha\alpha}(\theta)  & \bm{\rho}_{\alpha\beta}(\theta)   \\ 
                           \bm{\rho}_{\beta\alpha}(\theta)   & \bm{\rho}_{\beta\beta}(\theta)   \end{pmatrix}
\\
  &= \begin{pmatrix} \mathbf{P}(\theta) + \mathbf{M}_z(\theta)              & \mathbf{M}_x(\theta) - \mathrm{i} \mathbf{M}_y(\theta)   \\ 
                     \mathbf{M}_x(\theta) + \mathrm{i} \mathbf{M}_y(\theta) & \mathbf{P}(\theta) - \mathbf{M}_z(\theta)                \end{pmatrix}
\nonumber
\\
\bm{\kappa}(\theta)
  &\equiv  \begin{pmatrix} \bm{\kappa}_{\alpha\alpha}(\theta)  & \bm{\kappa}_{\alpha\beta}(\theta)   \\ 
                           \bm{\kappa}_{\beta\alpha}(\theta)   & \bm{\kappa}_{\beta\beta}(\theta)    \end{pmatrix}
\\
  &= \begin{pmatrix} -\bm{\kappa}_x(\theta) + \mathrm{i} \bm{\kappa}_y(\theta)  &  \bm{\kappa}_z(\theta) + \bm{\kappa}_0(\theta)             \\
                      \bm{\kappa}_z(\theta) - \bm{\kappa}_0(\theta)             &  \bm{\kappa}_x(\theta) + \mathrm{i} \bm{\kappa}_y(\theta)  \end{pmatrix}.
\nonumber
\end{align}

\section{Form of the Reduced Density Matrices}
To compute a general expectation value, it is most convenient to work with reduced density matrices.  We discuss here how to evaluate these tensors using the projected HFB state.  The $N$-particle reduced density matrix ($N$-RDM) corresponding to the projected HFB state is given by
\begin{align}
\Gamma_{i_1\cdots i_N,j_1 \cdots j_N} 
  &= \frac{1}{N!} \langle a_{j_1}^{\dagger} \cdots a_{j_N}^{\dagger} a_{i_N} \cdots a_{i_1} \rangle
\\
  &= \frac{\langle \Phi | \hat{P}^\dagger \, \hat{\Gamma}_{i_1 \cdots i_N,j_1 \cdots j_N} \, \hat{P} | \Phi \rangle}{\langle \Phi | \hat{P}^\dagger \, \hat{P} | \Phi\rangle}
\end{align}
where we have defined
\begin{equation}
\hat{\Gamma}_{i_1 \cdots i_N,j_1 \cdots j_N} = \frac{1}{N!} a_{j_1}^{\dagger} \cdots a_{j_N}^{\dagger} a_{i_N} \cdots a_{i_1}.
\end{equation}

Unfortunately, the operators $\hat{\Gamma}$ do not commute with the projection operators in general.  However, one can evaluate the reduced density matrices by double integration over the gauge angle.  That is,
\begin{equation}
\bm{\Gamma} = \frac{\iint \mathrm{d}\theta \, \mathrm{d}\theta' \, w(\theta) \, w(\theta') \, \langle \Phi | \hat{R}^\dagger(\theta') \, \hat{\Gamma} \, \hat{R}(\theta) | \Phi \rangle}{\iint \mathrm{d}\theta \, \mathrm{d}\theta' \, w(\theta) \, w(\theta') \, \langle \Phi | \hat{R}^\dagger(\theta') \, \hat{R}(\theta) | \Phi \rangle}.
\label{Npdm}
\end{equation}
Defining the normalized weighting function
\begin{equation}
y(\theta,\theta') = \frac{w(\theta) \, w(\theta') \langle \Phi | \hat{R}^\dagger(\theta') \, \hat{R}(\theta) | \Phi \rangle}{\iint \mathrm{d}\phi \, \mathrm{d}\phi' \, w(\phi) \, w(\phi') \langle \Phi | \hat{R}^\dagger(\phi) \, \hat{R}(\phi) | \Phi \rangle},
\end{equation}
the reduced density matrix becomes
\begin{equation}
\bm{\Gamma} = \iint \mathrm{d}\theta \, \mathrm{d}\theta' \, y(\theta,\theta') \langle \theta' | \hat{\Gamma} | \theta \rangle.
\end{equation}

The overlap matrix elements can be computed from
\begin{equation}
\langle \Phi | \hat{R}^\dagger(\theta') \, \hat{R}(\theta) | \Phi \rangle = \frac{\mathrm{det} \, \mathbf{R}_\theta \, \, \mathrm{det} \, \mathbf{R}_{\theta'}^\dagger}{\sqrt{\mathrm{det} \, \bm{\rho}} \sqrt{\mathrm{det} \, \mathbf{C}_{\theta \theta'}}}
\end{equation}
with
\begin{equation}
\mathbf{C}_{\theta \theta'}^{-1} =  
  \mathbf{R}_{\theta'} \, \bm{\rho} \, \mathbf{R}_{\theta'}^{\dagger} \, \mathbf{R}_\theta \, \bm{\rho} \, \mathbf{R}_\theta^\dagger
- \mathbf{R}_{\theta'} \, \bm{\kappa} \, \mathbf{R}_{\theta'}^\mathsf{T} \, \mathbf{R}_\theta^\star \, \bm{\kappa}^\star \, \mathbf{R}_\theta^\dagger.
\end{equation}
The transition density matrices needed to evaluate the density operator expectation values are in turn given by
\begin{align}
\bm{\rho}(\theta,\theta') &= \mathbf{R}_\theta \, \bm{\rho} \, \mathbf{R}_\theta^\dagger \, \mathbf{C}_{\theta\theta'} \, \mathbf{R}_{\theta'} \, \bm{\rho} \, \mathbf{R}_{\theta'}^\dagger,
\\
\bm{\kappa}(\theta,\theta') &= \mathbf{R}_\theta \, \bm{\rho} \, \mathbf{R}_\theta^\dagger \, \mathbf{C}_{\theta\theta'} \, \mathbf{R}_{\theta'} \, \bm{\kappa} \, \mathbf{R}_{\theta'}^\mathsf{T},
\\
\bar{\bm{\kappa}}^\star(\theta,\theta') &= \mathbf{R}_\theta^\star \, \bm{\kappa}^\star \, \mathbf{R}_\theta^\dagger \, \mathbf{C}_{\theta\theta'} \, \mathbf{R}_{\theta'} \, \bm{\rho} \, \mathbf{R}_{\theta'}^\dagger.
\end{align}
The $N$-PDM is then formed by integrating products of transition density matrices.  For example, the 2-PDM is
\begin{align}
\Gamma_{kl,ij} = \frac{1}{2} \iint \mathrm{d}\theta \, \mathrm{d}\theta' \, y(\theta,\theta')
   &\Big\{ \rho_{ki}(\theta,\theta') \rho_{lj}(\theta,\theta')
\\
   &- \rho_{kj}(\theta,\theta') \rho_{li}(\theta,\theta')
\nonumber
\\
   & + \bar{\kappa}^\star_{ij}(\theta,\theta') \kappa_{kl}(\theta,\theta')\Big\}
\nonumber
\end{align}
with similar factorizable expressions for all higher order density matrices.


%
\end{document}